\documentclass[a4paper,fleqn,usenatbib]{mnras}
\pdfoutput=1

\usepackage{newtxtext,newtxmath}

\usepackage[T1]{fontenc}
\usepackage{ae,aecompl}

\usepackage{graphicx}	
\usepackage{amsmath}	
\usepackage{dblfloatfix}
\usepackage{ulem}

\usepackage{xspace} 

\newcommand{\de}{\ensuremath{\mathrm{d}}}

\usepackage[dvipsnames]{xcolor}

\title[SPIDERS Galaxy Cluster Catalogue]{SPIDERS: An Overview of The Largest Catalogue of Spectroscopically Confirmed X-ray Galaxy Clusters}

\author[C. C. Kirkpatrick et al.]{
C.~C.~Kirkpatrick,$^{1,2}$\thanks{E-mail: charles.kirkpatrick@helsinki.fi}
N.~Clerc,$^{3}$
A.~Finoguenov,$^{1}$
S.~Damsted,$^{1}$
J.~Ider~Chitham,$^{4}$
\newauthor
A.~E.~Kukkola,$^{1}$
A.~Gueguen,$^{4}$
K.~Furnell,$^{5}$
E.~Rykoff,$^{6,7}$
J.~Comparat,$^{4}$
A.~Saro,$^{8,9,10}$
\newauthor
R.~Capasso,$^{11}$
N.~Padilla,$^{12}$
G.~Erfanianfar,$^{4}$
G.~A.~Mamon,$^{13}$
C.~Collins,$^{5}$
A.~Merloni,$^{4}$
\newauthor
J.~R.~Brownstein,$^{14}$
and D.~P.~Schneider$^{15,16}$
\\
$^{1}$Department of Physics, University of Helsinki, Gustaf H{\"a}llstr{\"o}min katu 2, FI-00014 Helsinki, Finland \\
$^{2}$Helsinki Institute of Physics, Gustaf H{\"a}llstr{\"o}min katu 2, FI-00014 Helsinki, Finland \\
$^{3}$IRAP, Universit{\'e} de Toulouse, CNRS, UPS, CNES, Toulouse, France \\
$^{4}$Max-Planck-Institut f{\"u}r extraterrestrische Physik, Giessenbachstrae, 85748 Garching, Germany \\
$^{5}$Astrophyics Research Institute, Liverpool John Moores University, IC2, Liverpool Science Park, 146 Brownlow Hill, Liverpool L3 5RF, UK \\
$^{6}$Kavli Institute for Particle Astrophysics \& Cosmology, P. O. Box 2450, Stanford University, Stanford, CA 94305, USA \\
$^{7}$SLAC National Accelerator Laboratory, Menlo Park, CA 94025, USA \\
$^{8}$Astronomy Unit, Department of Physics, University of Trieste, via Tiepolo 11, I-34131 Trieste, Italy \\
$^{9}$IFPU - Institute for Fundamental Physics of the Universe, Via Beirut 2, 34014 Trieste, Italy \\
$^{10}$INAF-Osservatorio Astronomico di Trieste, via G. B. Tiepolo 11, I-34143 Trieste, Italy \\
$^{11}$The Oskar Klein Centre, Department of Physics, Stockholm University, Albanova University Center, SE 106 91 Stockholm, Sweden \\
$^{12}$Instituto de Astrof{\'i}sica, Pontificia Universidad Cat{\'o}lica de Chile, Av. Vicuna Mackenna 4860, 782-0436 Macul, Santiago, Chile \\
$^{13}$Institut d'Astrophysique de Paris, (UMR 7095: CNRS~$\And$~UPMC, Sorbonne Universit{\'e}), F-75014 Paris, France \\
$^{14}$Department of Physics and Astronomy, University of Utah, 115 S. 1400 E., Salt Lake City, UT 84112, USA \\
$^{15}$Department of Astronomy and Astrophysics, The Pennsylvania State University, University Park, PA 16802 \\
$^{16}$Institute for Gravitation and the Cosmos, The Pennsylvania State University, University Park, PA 16802
}

\date{Accepted XXX. Received YYY; in original form ZZZ}

\pubyear{2019}

\begin{document}
\label{firstpage}
\pagerange{\pageref{firstpage}--\pageref{lastpage}}
\maketitle

\begin{abstract}
SPIDERS is the spectroscopic follow-up effort of the Sloan Digital Sky Survey IV (SDSS-IV) project for the identification of X-ray selected galaxy clusters.  We present our catalogue of 2740 visually inspected galaxy clusters as a part of the SDSS Data Release 16 (DR16).  Here we detail the target selection, our methods for validation of the candidate clusters, performance of the survey, the construction of the final sample, and a full description of what is found in the catalogue.  Of the sample, the median number of members per cluster is approximately 10, with 818 having 15 or greater.  We find that we are capable of validating over 99\% of clusters when 5 redshifts are obtained below $z<0.3$ and when 9 redshifts are obtained above $z>0.3$.  We discuss the improvements of this catalogue's identification of cluster using 33,340 redshifts, with $\Delta z_{\rm{phot}} / \Delta z_{\rm{spec}} \sim 100$, over other photometric and spectroscopic surveys, as well as present an update to previous ($\sigma - L_{X}$) and ($\sigma - \lambda$) relations.  Finally, we present our cosmological constraints derived using the velocity dispersion function.
\end{abstract}




\section{Introduction}

Galaxy clusters are the most massive bound objects in the Universe.  The more massive, $10^{14} - 10^{15}$ M$_{\odot}$, the more rare.  As a trace of the underlying cosmic web, they can be used as an independent probe of the expansion rate of the Universe and growth of structure.  Well measured populations can place constraints on cosmological models based on their predictions \citep[e.g.][]{boh04,vik09,man10,wei13}.  Clusters are X-ray luminous sources due to the large amount of hot gas trapped in their gravitational potential wells \citep{jon99}.  The extended nature of the sources make them distinct from most other extra-galactic sources, allowing more ease in performing surveys of high volume and completeness.  Additionally, X-ray properties of clusters scale with the mass of the system \citep{kai86}, making these types of surveys even more important for cosmological studies.  

The evolution of the halo mass function is an important cosmological tool. Clusters are best detected from the X-rays emitted by their hot diffuse gas.  The first major X-ray cluster survey came with the ROSAT All Sky Survey \citep{ebe00,ike02,rei02,sch03}, mapping the brightest galaxy clusters outside of the Galactic Plane.  Subsequent surveys \citep{ros98,rom00,burk03,bure07} and other generations of X-ray telescopes have contributed to cluster cosmology \citep{pac06,vik09,man10,fin10,cle14,pie16}, though not on an all-sky scale.  The next advance in X-ray astronomy is the newly launched extended ROentgen Survey with an Imaging Telescope Array (eROSITA).  Performing a new all-sky survey in the soft X-ray band, the improved resolution and sensitivity will lead to $\sim$100\,000 galaxy cluster detections ($\sim3 \times 10^{-14}$ erg s$^{-1}$ cm$^{-2}$ flux limit at 0.5-2 keV) out to a redshift of $\sim$1 \citep{mer12,bor14}.

Though X-rays are most efficient for the purpose of cluster detection and mass proxy, the understanding of their redshift to high accuracy is lacking.  This highlights the importance of optical follow-ups.  The Sloan Digital Sky Survey (SDSS) has a played a critical role in optical confirmation of galaxy clusters.  Measuring the red-sequence of galaxy cluster member galaxies using multi-waveband optical imaging \citep{gla00,ryk14} provides a useful tool for identifying the optical counterparts of clusters and characterising their redshifts.  For the ultimate purpose of constraining cosmological models, spectroscopic redshifts are needed.  The extra precision on redshift allows for the ability to disentangle projection effects of line-of-sight alignments, infalling substructure, and other correlated structure from the underlying dark matter halo concentration.  The next important step is the difficult task of obtaining galaxy spectra, which requires deep imaging for targeting and longer spectroscopic integration times.

The SPectroscopic IDentification of eROSITA Sources (SPIDERS) cluster program within SDSS-IV \citep{bla17} was designed to carry out the necessary follow-up for the next generation of all-sky X-ray surveys.  The survey itself fully relies on the use of the Baryon Oscillation Spectroscopic Survey (BOSS) spectrograph \citep{sme13} mounted on the SDSS-2.5m telescope at Apache Point Observatory \citep{gun06}.  Our observing strategy is shared with the cosmology project eBOSS (extended Baryon OScillation Survey, \citet{daw16}) in order to maximize the ability to obtain large amounts of spectroscopic redshifts.  SPIDERS targets are few in comparison to eBOSS, but are given the highest priority when assigning fibers for observing.  

An initial study of the first 300 deg$^2$ of the survey were provided by \citet{cle16}.  A small sample of galaxy clusters along with the full target list were published in the form of a Value Added Catalogue\footnote{\texttt{https://www.sdss.org/dr14/data\_access/value-added-catalogs/?vac\_id=the-spiders-clusters-demonstration-sample-catalogue}} (VAC).  This study demonstrated the survey's feasibility and usefulness of the data set.  Additional studies have been preformed using an updated VAC catalogue\footnote{\texttt{https://www.sdss.org/dr14/data\_access/value-added-catalogs/?vac\_id=spiders-x-ray-galaxy-cluster-catalogue-for-dr14}} as a part of Data Release 14 (DR14).  This has been used for topics including characterising scaling relations with cluster richness between both mass and X-ray luminosity \citep{cap19,cap20} and properties of brightest cluster galaxies \citep{fur18,erf19}.

The COnstraining Dark Energy with X-ray (CODEX) clusters survey \citep{fin20} describes the X-ray source  catalogue that forms the basis for the SPIDERS follow-up program.  They present the advanced wavelet filtering techniques used on X-ray images and provides the modeling of the sample selection.  A companion paper to this work \citep[Clerc et al. accepted][]{} describes in detail the final status of the survey including: overall characteristics, final targeting strategies, achieved completeness and spectral quality.  Cosmological applications of the galaxy cluster sample are explored.

This paper describes the entire process for the construction of the final SPIDERS galaxy cluster catalogue as part of DR16 \citep{ahu20}.  It encompasses all data obtained throughout the entire SDSS-IV program, including the original pilot sample of \citet{cle16}.  Every aspect of the process is detailed from target selection, manual inspection of the spectroscopic members and analysis of the full data set.  We present an update to the first results from the SPIDERS cluster program.

The outline of this paper is as follows.  In Section~\ref{sec:2} we describe the SPIDERS sample and how targets were selected for the survey.  In Section~\ref{sec:3} we describe our method of visual inspection for validating the spectroscopic redshift of the cluster.  In Section~\ref{sec:4} we present the final results of the validation effort.  Section~\ref{sec:5} details the construction of what goes into the final catalogue.  Section~\ref{sec:6} is the full description of the value added catalogue with highlights of its scientific importance.  Finally, in Section~\ref{sec:7} we present our initial cosmological constraints based on our velocity dispersion measurements.

Unless otherwise stated, the cosmological model used in this paper is a flat $\Lambda$ Cold Dark Matter with $\Omega_{\rm{m}} = 0.3$ and $H_{\rm{o}}= 70$ km s$^{-1}$ Mpc$^{-1}$. Magnitudes are expressed in their native SDSS (AB) system \citep{fuk96}.

\section{The SPIDERS Cluster Sample} \label{sec:2}

Cluster candidates for SPIDERS have been drawn from a subset of CODEX \citep{fin20}.  CODEX is an X-ray selected catalogue of clusters reaching fluxes of 10$^{-13}$ ergs s$^{-1}$ cm$^{-2}$.  CODEX uses data from the ROSAT All Sky survey (RASS), specifically in the $\sim$10,000 square degree BOSS imaging footprint.  A comprehensive description of source detection is presented in \citet{fin20}.  A total of 10,415 CODEX entries were considered for follow-up, of which 4,114 made it into the final area surveyed by SPIDERS.

Target galaxies for spectroscopic follow-up are identified using the red-sequence technique.  For each X-ray source detection, redMaPPer version 5.2 \citep{ryk14} was run at that position on the sky using the SDSS imaging data of  Data Release 8 (DR8). The calibration of the red-sequence is carried out using existing SDSS spectroscopy from DR9 and only needs a small set of training clusters for calibration.  The full details of this procedure is presented in \citet{ryk14}.  Given a new red-sequence model, cluster finding begins by looking for galaxy members in a 0.7 Mpc radius of the X-ray centre.  The algorithm proceeds to iterate over each guess for the redshift.  An optimized richness estimator \citep{roz09,ryk12} uses colour offset from the red-sequence, $i$-band magnitude, and projected distance from the cluster centre to determine the probability of a galaxy being a member, ultimately producing a richness and it's corresponding likelihood.  Richness is defined as the sum of the probabilities, with corrections for missing area and photometric depth (see \citet{ryk14} for full details).  If more than 3 members are detected, the highest probability members are simultaneously fit to the red-sequence model to find the new cluster photometric redshift.  This process iterates until convergence on redshift ($|z_{i+1} - z_{i}| < 0.0002$) is achieved.  The candidate cluster with the highest likelihood at the end of this procedure is selected as the optical counterpart.  

Due to the uncertainty associated with the position of RASS detections, the entire algorithm is run again allowing for the centre to vary within 3 arcmin.  This new red-sequence provides an alternate cluster centre and the final estimate of the cluster's photometric redshift ($z_{\rm{\lambda}}$) and richness ($\lambda_{\rm{OPT}}$) that are entered into the SPIDERS cluster sample.  We refer to this as the `optical detection'.

The reproducible process of associating initial red-sequence candidates to spectroscopic fibres is referred to as 'targeting'. It involves all sub-components of the eBOSS programme. It is designed to optimize efficiency metrics defined for each survey; in the case of SPIDERS, the figure of merit is spectroscopic completeness. The targeting algorithm uses priority flags associated to each targets. We use the galaxy membership probability as the main factor in determining SPIDERS clusters targeting priority.  Before ranking, galaxies that we do not want to be targeted are filtered out.  All galaxies associated with clusters of richness less than 10 are removed from the target list.  All galaxies with a probability less than 5 percent are also removed.  Members already with a spectroscopic redshift from past observations are matched and removed.  For the remaining galaxies, only members with a fiber i-band magnitude between 17.0 and 21.2 will be considered for targeting to maximize detection efficiency. These filtering steps led from an initial list of 158,368 CODEX red-sequence galaxies to a pool of 44,367 galaxies submitted for follow-up. This defines the full SPIDERS member candidate sample.

The final target ranking is determined, in general, as a function of cluster richness and member probability.  Cluster richness is divided into bins of: $\lambda_{\rm{OPT}} > 40$, $30 < \lambda_{\rm{OPT}} < 40$, $20 < \lambda_{\rm{OPT}} < 30$, $10 < \lambda_{\rm{OPT}} < 20$.  Each cluster's red-sequence is ranked from highest to lowest probability.  The three top ranked members of all clusters are assigned the same priority flag of 1, 2, and 3, regardless of richness.  From rank 4 and up, priority is given to galaxies in higher richness bins.  After this ranking, a subsequent filtering occurs due to physical conditions during observing: sky fiber density is restricted to $\sim$10/deg$^2$ and fiber collisions at a radius of 62'' within an SDSS plate. Finally, 30,236 galaxies were selected for follow-up within the DR16 sky area.  The full details and visualisation of this scheme are found in \citet{cle16} and more recent updates reported in \citet{cle20}.

\section{Spectroscopic Validation} \label{sec:3}

Trained astronomers are required to verify the existence of a cluster in order to disentangle their complex nature \citep{guz09, ada11}.  An automatic membership assignment algorithm was developed to ease the validation efforts of such a large sample \citep{cle16}.  This was implemented as the first step before any visual inspection takes place.  Our approach is broad in order to account for varying mass throughout the sample.  Each cluster with red-sequence members associated with a spectroscopic redshift are considered individually.  An initial redshift is computed using the bi-weight average \citep[see,][]{bee90} of all observed members, $N_{\rm{zspec,0}}$.  All members are rejected with velocities offset by greater then 5000 km s$^{-1}$.  A $3\sigma$ clipping procedure is run on the remaining members, $N_{\rm{zspec,1}}$, to converge on a final preliminary redshift and velocity dispersion.  The procedure starts with recomputing the bi-weight average, $z$, of the remaining members.  The proper velocity for each galaxy is estimated by $v_{\rm{prop}} = c(z_i - z) / (1 + z)$ \citep{dan80}, where $z_i$ is the redshift of the individual galaxy.  Velocity dispersion is then calculated depending on the number of spectroscopic members.  For cases with $N_{\rm{zspec,1}} \geq 15$, velocity dispersion is calculated as the square root of the bi-weight variance of the member galaxies' proper velocities.  Otherwise, the gapper method is used for cases with a lower spectroscopic sampling rate, as it is known to give more robust results \citep{bee90}.  The entire procedure is allowed to iterate until convergence or is stopped after 10 steps.

\begin{figure}
    \centering
    \includegraphics[width=\columnwidth]{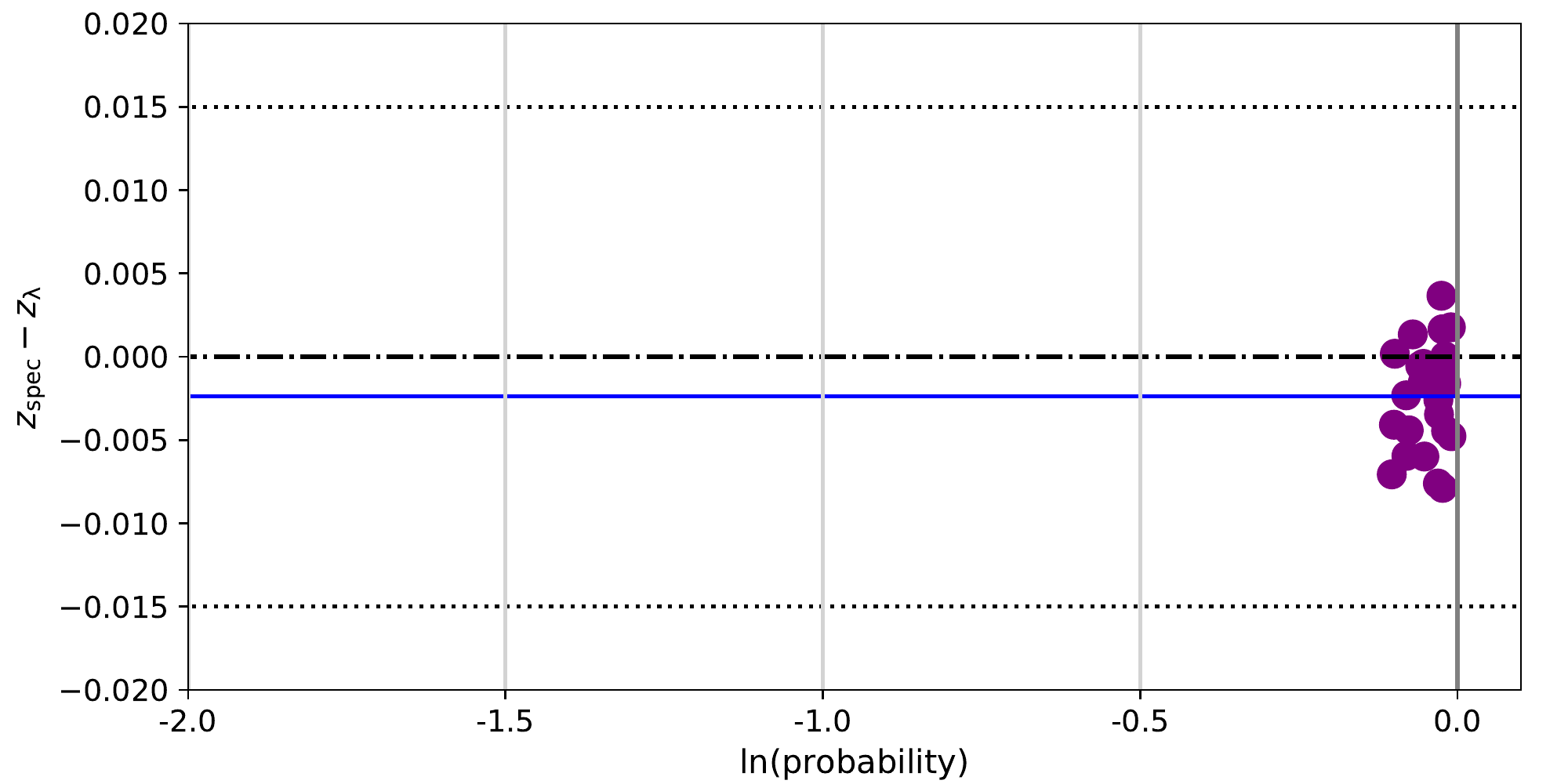}
    \includegraphics[width=\columnwidth]{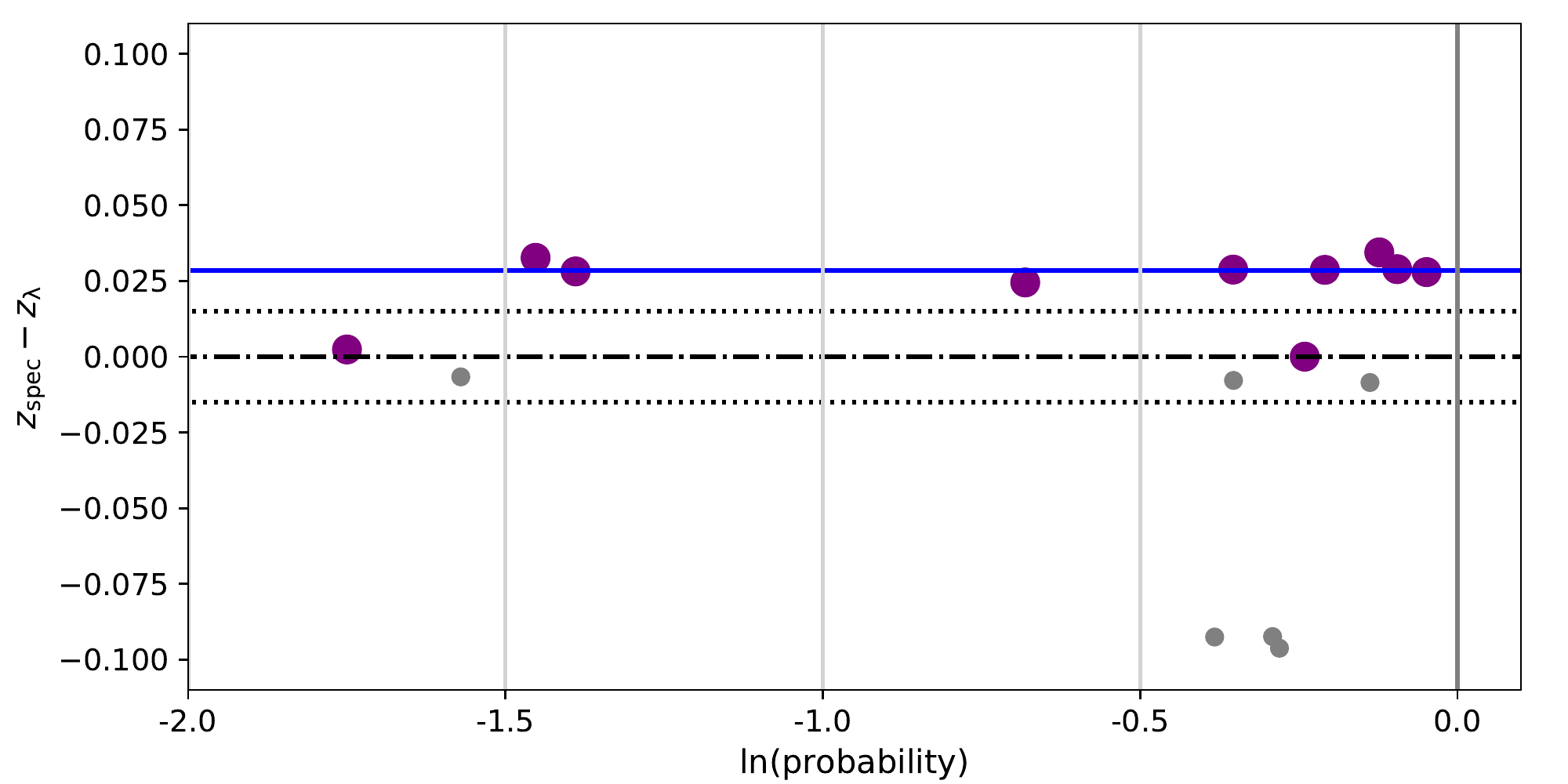}
    \caption{Automatic analysis of redshift (velocities) for two different clusters.  Plotted are the difference between spectroscopic and cluster photometric redshift versus RedMaPPer membership probability.  Points highlighted in purple were selected as members by the automatic algorithm and the blue line is the cluster redshift determined by that selection.  The dash-dot and dotted lines are the photometric redshift and it's associated error.  Upper panel: This example shows a cluster where the automatic assignment has no problem identifying based solely on the $3\sigma$ clipping procedure.  Lower panel: Another example where automatic membership assignment fails due to confusion with possible nearby structure.  The grey points are galaxies not selected as members by the automatic algorithm.  These figures are representative of diagnostic plots used during inspection.}
    \label{fig:automatic_result}
\end{figure}

Figure~\ref{fig:automatic_result} highlights two cases, one where the automatic membership assignment works successfully and one where it fails.  The upper panel of this figure is the typical example where no interference is required during visual inspection.  Little time is needed in a case like this when the cluster is well sampled and no signs of outliers are present.  The lower panel shows where the automatic assignment has trouble when line-of-sight contamination is strong.  A combination of studying the phase space diagrams, 2D projected maps, velocity histograms, optical images, and X-ray detection information is needed to understand and recalculate membership.  The details of what is available during visual inspection is described below.

\begin{figure*}
    \centering
    \includegraphics[width=\textwidth]{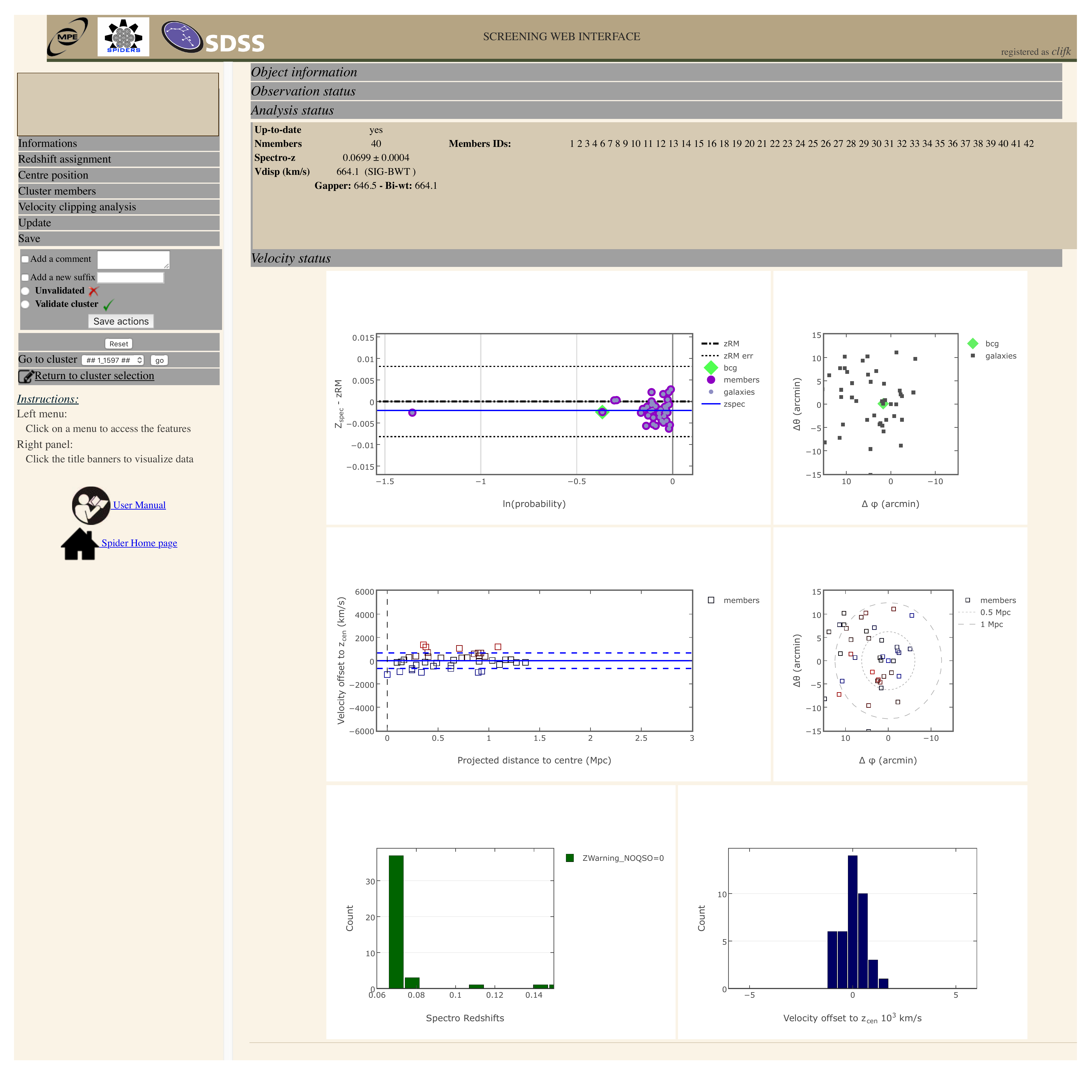}
    \caption{Screening Web Interface.  Various phase-space plots are used to help inspectors make their decision whether or not to validate a cluster.
    {/bf Red-sequence galaxies with a spectroscopic redshift are represented in the uppermost panels, as a function of their redshift offset to the cluster photometric redshift (on the left) of their position on sky (on the right).}
    Members are colour coded by negative (blue) or positive (red) proper velocity in the middle-left and middle-right panel, showing respectively the velocity-distance distribution and sky distribution of current selected members. The lower left histogram shows the redshift distribution of all red-sequence galaxies. The lower right histogram stands for current selected members only. Tools for manually assigning members for recalculating redshift and velocity dispersion are also available from the command panels on the left-hand side of the interface.}
    \label{fig:screening}
\end{figure*}

Figure~\ref{fig:screening} shows a typical snapshot of the screening web interface used by the visual inspectors. The six panels within the interface are as follow: the top-left panel is the difference between spectroscopic redshift and photometric redshift plotted against the natural log of the redMaPPer membership probability for all observed red-sequence members. Red-sequence galaxies selected as cluster members initially by the algorithm (or modified later by the user) are highlighted in purple, otherwise they would be shown as grey points. The top-right panel is the 2D projected sky positions of all observed red-sequence members, the middle-left panel is the velocity offset to the computed cluster redshift plotted against the projected distance from the cluster center in Mpc for only the selected cluster members, the middle-right panel is the 2D projected sky positions of only the selected members with the physical distance from the cluster center highlighted in 0.5 Mpc intervals, the bottom-left panel is the distribution of red-sequence member redshifts, and the bottom-right panel is the velocity offset distribution of only the selected cluster members.

Catalogue information is available to the inspectors in the upper tabs of the interface.  They contain, most importantly for inspection, the photometric redshift, the richness estimate, and how many red-sequence members have spectroscopy.  The tabs on the left side allow the inspectors to access more information about the cluster and utilize tools to aid in the manual selection of members in cases where the automatic algorithm fails.  During validation, inspectors have access to all the individual spectra, direct links to the cluster location in legacy sky viewer \citep{dey19}, and visual displays of all SDSS related spectroscopy in the field of view together with all red-sequence members.  Different tools allow the inspector to manually select or de-select single or groups of members.  Redshift and velocity dispersion is always recalculated using the same routines used by the automatic algorithm.

During validation, no requirements were given to the inspectors in order to ensure independence between visual inspections.  Loose guidelines were discussed amongst inspectors, mainly concerning available literature on previous work. Studies of nearby systems \citep{rin13, mun14} and scaling relations based on richness \citep{sim17,mur18} where important for understanding the make-up of galaxy clusters. The former benefit from galaxy phase-space diagrams sampled by factors several tens of hundreds denser than typical SPIDERS clusters; these clusters were considered are high signal-to-noise analogs. The latter provide a support for excluding configurations incompatible with the expected scaling of cluster observables with total mass (e.g.~low-richness clusters should have small velocity dispersions and conversely).

\begin{figure*}
    \centering
    \includegraphics[width=\textwidth]{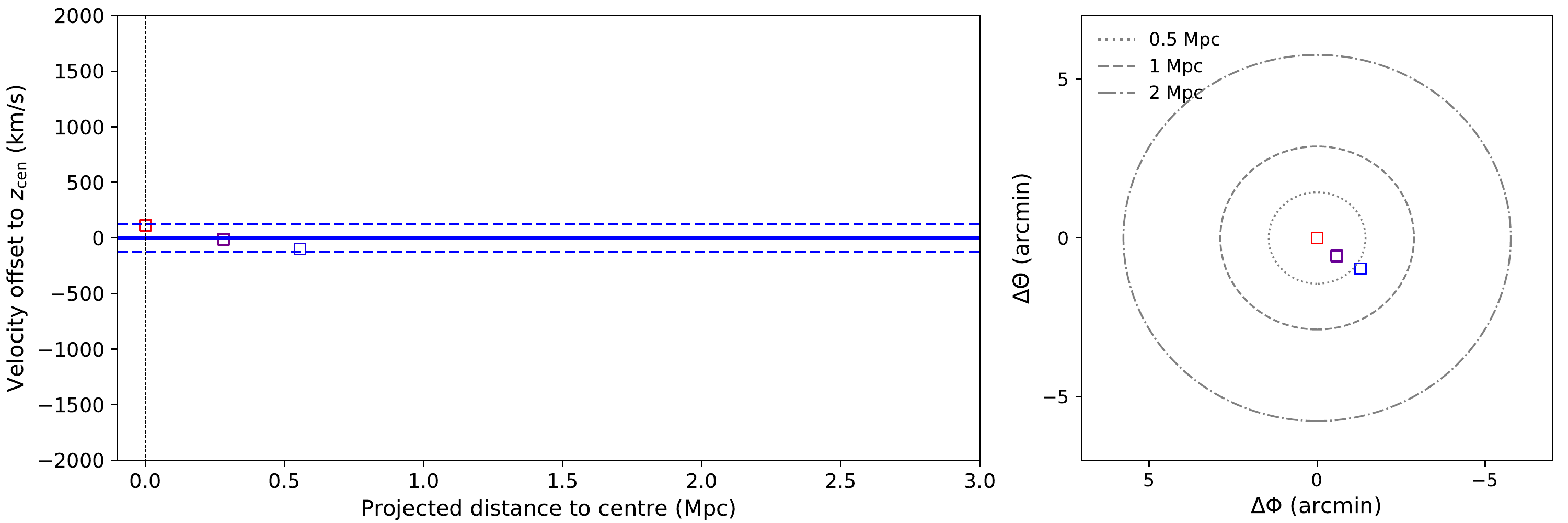}
    \includegraphics[width=\textwidth]{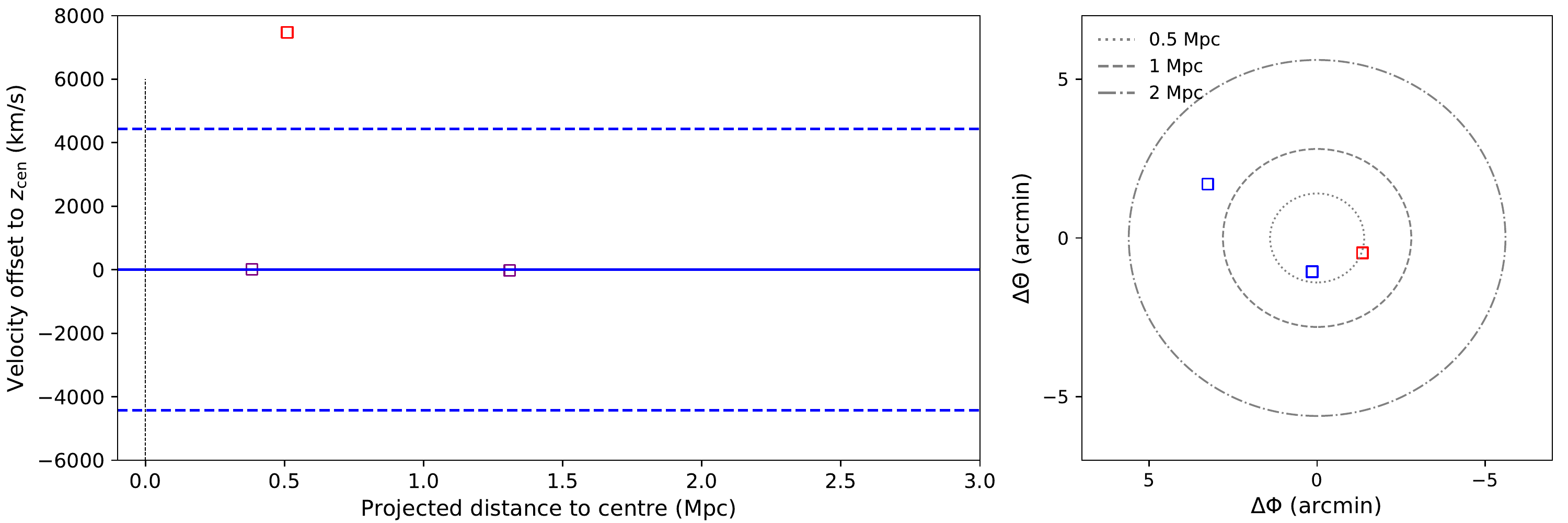}
    \caption{Illustration of two clusters with only 3 candidate member galaxies.  A blue-to-red colour gradient is used to highlight the negative-to-positive velocity offset from the centre.  Upper panels: Velocity offset and distance phase space plot with the 2D sky projecting from the screening web tool.  This example illustrates the case when the minimum number of red-sequence members are able to validate the cluster redshift. Lower panels: Same plots as above, but for a case where validation cannot be confirmed, as not even two of galaxies appear to likely be associated with one another.  Solid and dashed lines are the same as shown in Figure~\ref{fig:automatic_result}.}
    \label{fig:3mem}
\end{figure*}

To further illustrate the validation procedure, we will highlight a few examples of common scenarios encountered during the process.  A minimum of three members are required to converge on a final redshift solution. These members should lie close to each other in the velocity-distance diagram. Explicit formulations of exact threshold criteria was avoided and everything was left to inspectors as much as possible; such criteria emerged in the later stage of reconciliation among inspectors (as described below). In Figure~\ref{fig:3mem}, the upper panels shows a case where three members are sufficient to validate a redshift.  In the velocity-distance phase space plot, all three members are at similar velocities, and are also clustered in 1D distance and in 2D sky projection.  Coinciding with an X-ray detection as well gives us confidence this is more than a chance pairing.  In the lower panels is a case where three observations is not sufficient to validate a redshift.  One galaxy is offset in velocity from the other two members by more than 6000 km s$^{-1}$, while these same two members are separated on the sky by over a megaparsec.  More observations would be needed to determine which of these three, if any, are a part of a cluster at this location.

\begin{figure}
    \centering
    \includegraphics*[width=\columnwidth]{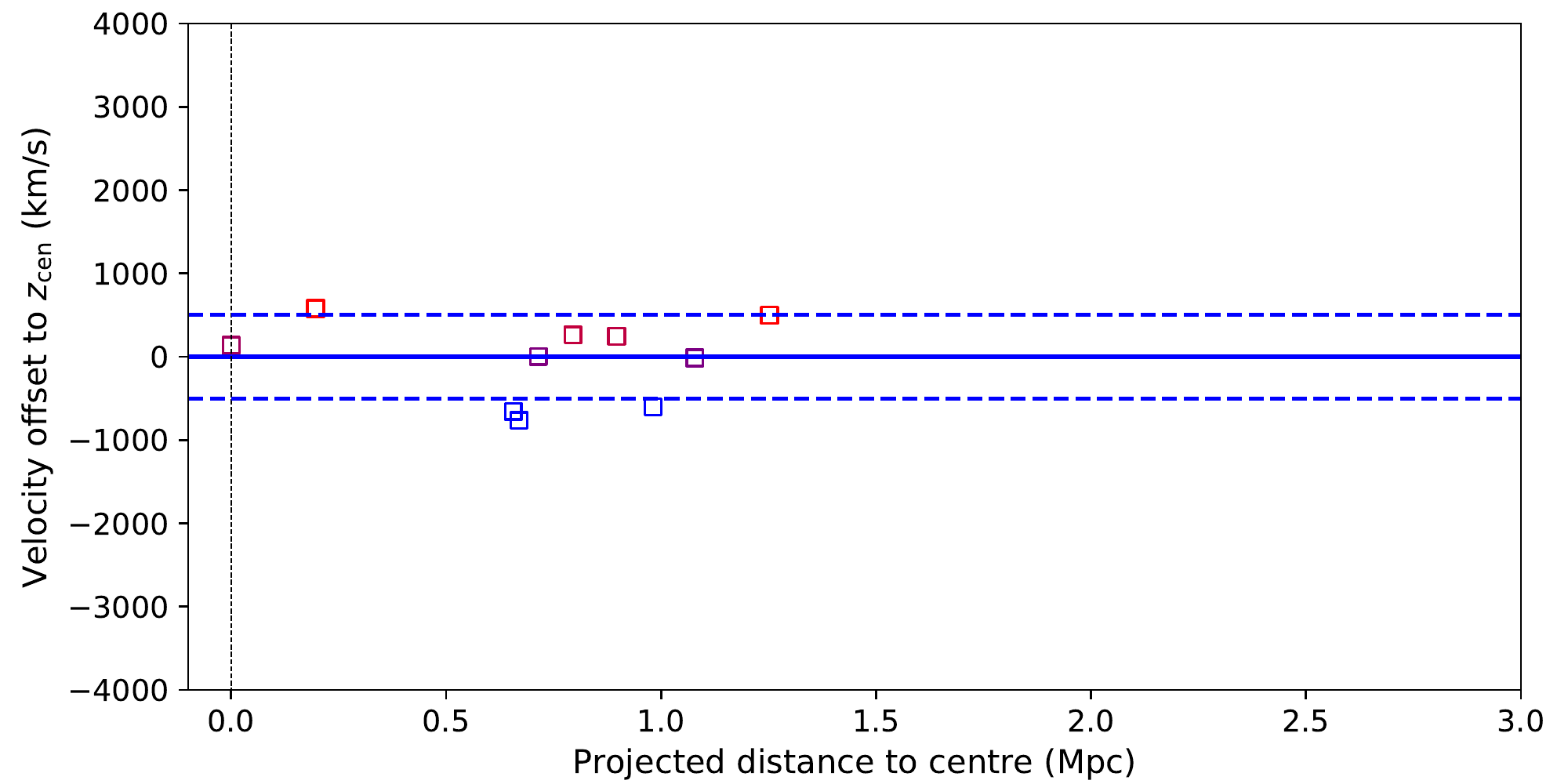}
    \includegraphics*[width=\columnwidth]{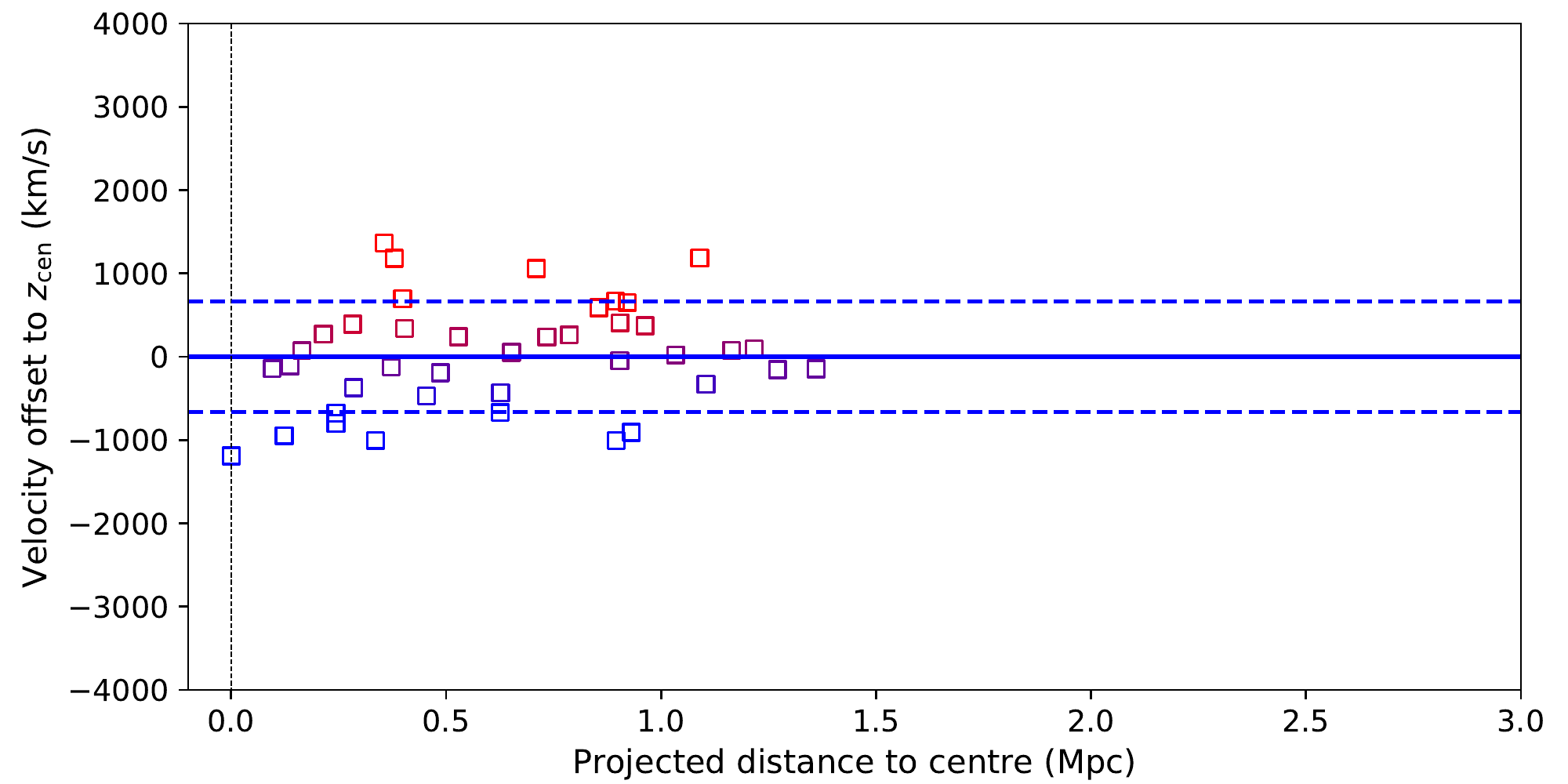}
    \caption{Illustration of clusters with intermediate and high number of candidate member galaxies.  The points are highlighted using the same colour scheme of Figure~\ref{fig:3mem}.  Upper panel: Velocity offset and distance phase space plot for a cluster with the number of members equal to the median.  Lower panel: The same phase space plot for a rare, well sampled system. Solid and dashed lines are the same as shown in Figure~\ref{fig:automatic_result}.}
    \label{fig:ave_clus}
\end{figure}

 An example of a cluster with 10 members, the median value after validation, is illustrated in the upper panel of Figure~\ref{fig:ave_clus}.  Universally these are agreed upon that they are in fact clusters, but on an inspector by inspector basis, there may not be agreement on the exact membership make up.  These differences are mediated in the reconciliation round of validation.  For comparison, the bottom panel of Figure~\ref{fig:ave_clus} shows the more rare instance of a well sampled system where it is easier to find a consensus, since agreement on clusters status and membership list are reached by all.

\begin{figure}
    \centering
    \includegraphics*[width=\columnwidth]{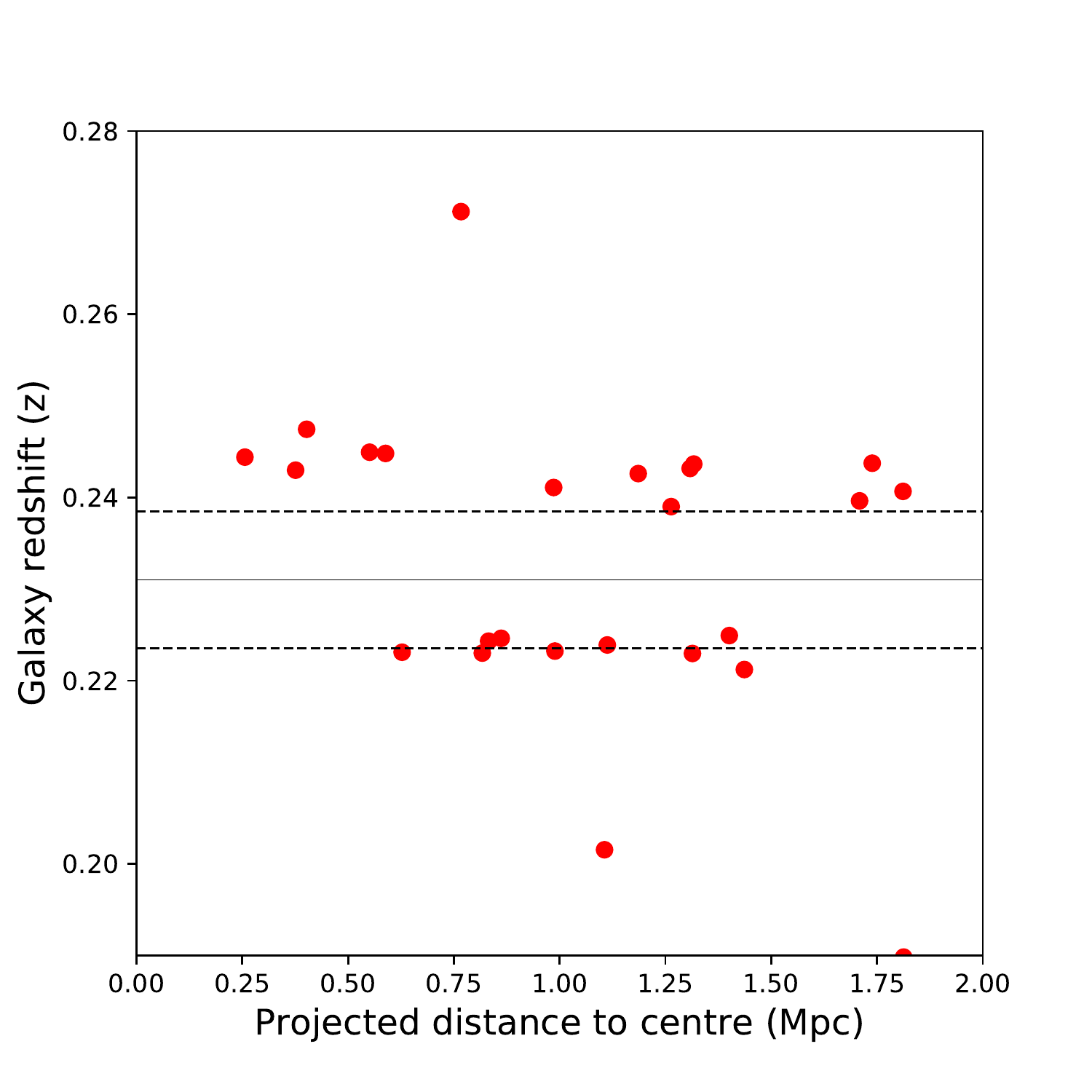}
    \caption{Redshift versus distance from cluster centre.  The solid line is the photometric redshift from redMaPPer and the dashed lines are the associated uncertainties on that estimation.  Two components are apparently at a lower and higher redshift compared to the photometric estimate are confusing the red-sequence finder algorithm.  They are likely similar in mass with a detection of 9 members in the lower redshift component and 13 members in the higher redshift component.  This results in an overestimation of richness by approximately a factor of two.}
    \label{fig:split_clus}
\end{figure}

The most time consuming aspect of validation belongs to the targets where more than one ``component'' is apparent along the line-of-sight. This affects $\sim 8\%$ of the total validated sample. We define a component as a grouping of 3 or more galaxies associated with an X-ray source.  There is no limit on how many component can be classified.  Upon the conclusion of the validation process, inspectors agreed on three main scenarios that were encountered: the redMaPPer redshift ($z_{\rm{\lambda}}$) corresponds to the main component, corresponds to a secondary component, or does not correspond to any component.  The first is the most common case we find representing approximately 50\% of all multi-component systems.  Multiple components can be apparent in the foreground or background, but the majority of observed galaxies are consistent with $z_{\rm{\lambda}}$, suggesting the performance of redMaPPer is mostly not affected by the presence of smaller systems, though the estimated richness has been overestimated by a small factor. The rarest case ($\sim 10\%$) is the second example.  A small grouping of galaxies is consistent with $z_{\rm{\lambda}}$, but a larger grouping is found in the foreground or background, outside the photometric errors.  Small components are defined to have less than half the number of members compared to the main component. We find the median red-sequence width before membership cleaning, $\Delta z$, to be a factor of 2 larger than the errors on $z_{\rm{\lambda}}$ for SPIDERS clusters.  This could be the possible explanation for why this offset of the main component occurs. The last case describes the rest of the systems, for which neither of the component's redshift are consistent with $z_{\rm{\lambda}}$, but the richness estimate is still accurate, or multiple components are confusing redMaPPer entirely.  Figure~\ref{fig:split_clus} presents an example of two equal components, split in between by $z_{\rm{\lambda}}$.  We find that when two, likely equal mass clusters are in near proximity to each other, redMaPPer interprets this as one cluster located between the real clusters with a richness overestimated by at least a factor of two.  We discuss in more detail how we define and report the main component of clusters below in section~\ref{sec:5}.

The final stage of validation is the reconciliation between the decisions from all inspectors. Individual evaluations are collected in the form of digital pre-formatted files at the end of an evaluation campaign. A moderator checks the usability of the submitted results and for each cluster candidate compares the output of all inspectors. This consists of a two-step process. The first step involves addressing the agreement between evaluations. It requires that a majority emerges among inspectors concerning the validation status of the cluster (either validated or unvalidated) and that redshift values agree within their 95\% uncertainty ranges. An automated algorithm attempts to find disjunct groups of redshift values close to each other. If a group emerges gathering a majority of inspections (or if all evaluations belong to one single group), the redshifts are said to agree with each other. In case of multiple component splits, the moderator is prompted for an explicit agreement. All candidates that do not reach an agreement are put back to the pool of inspections in order to collect more votes. The second step involves building a conciliated catalogue based on the agreed upon evaluations. If a majority of inspectors validated a candidate, the corresponding membership flags are averaged (column \texttt{SCREEN\_NMEMBERS\_W}) and so are the cluster systemic redshift, velocity dispersion and uncertainties. A redshift spread is computed as the standard deviation of the inspectors' cluster redshift values, reflecting the amplitude of the agreement between inspectors.  It is then added in quadrature to the cluster redshift uncertainty. In cases of multiple components, a manual association between components is made and averages are performed sequentially for each component. All unvalidated evaluations are discarded. Candidates collecting a majority of unvalidated statuses are flagged as such.
 
\begin{table*}
    \centering
    \begin{tabular*}{\textwidth}{c@{\extracolsep{\fill}}cccccr}
        \hline
        \multicolumn{1}{l}{ } & \multicolumn{1}{c}{Inspection} & \multicolumn{1}{c}{ } & \multicolumn{1}{c}{ } & \multicolumn{1}{c}{Validated } & \multicolumn{1}{c}{Unique} & \multicolumn{1}{r}{ } \\
        \multicolumn{1}{l}{Run} & \multicolumn{1}{c}{round} & \multicolumn{1}{c}{Selection} & \multicolumn{1}{c}{Candidates} & \multicolumn{1}{c}{components} & \multicolumn{1}{c}{components} & \multicolumn{1}{r}{Inspectors} \\
        \hline
        2016-07-04 & Round-1 & $\mathrm{\lambda}_\text{OPT}$ $>$ 30 & 573 & 520 & 478 & 8 \\
        2016-07-04 & Round-2 & 20 $<$ $\mathrm{\lambda}_\text{OPT}$ $<$ 30 & 344 & 246 & 230 & 8 \\
        2016-07-04 & Round-3 & 10 $<$ $\mathrm{\lambda}_\text{OPT}$ $<$ 20 & 618 & 396 & 344 & 2 \\
        2018-04-27 & Round-1 & $\mathrm{\lambda}_\text{OPT}$ $>$ 30 & 434 & 379 & 365 & 6 \\
        2018-04-27 & Round-2 & 20 $<$ $\mathrm{\lambda}_\text{OPT}$ $<$ 30 & 216 & 166 & 145 & 4 \\
        2018-04-27 & Round-3 & 10 $<$ $\mathrm{\lambda}_\text{OPT}$ $<$ 20 & 432 & 281 & 264 & 5 \\
        2018-04-27 & Round-4 & $\mathrm{\lambda}_\text{OPT}$ $>$ 10, incomplete (SEQUELS) & 193 & 175 & 159 & 2 \\
        2018-12-04 & Round-1 & $\mathrm{\lambda}_\text{OPT}$ $>$ 10, w/ revised redshift & 383 & 299 & 276 & 4 \\
        2019-03-22 & Round-1 & $\mathrm{\lambda}_\text{OPT}$ $>$ 10 & 818 & 616 & 571 & 3 \\
        2019-03-22 & Round-2 & $\mathrm{\lambda}_\text{OPT}$ $>$ 10, remaining partially observed & 140 & 101 & 87 & 3 \\
        2019-03-22 & Round-3 & re-inspection required & 14 & 14 & 14 & 2 \\
     \end{tabular*}
    \caption{Statistics for each round during the almost three year validation effort.  Subsequent runs are for the inspection of completed observations of candidates not covered in earlier runs, unless otherwise noted in the text.  }
    \label{tab:validation_results}
\end{table*}

\section{Validation results} \label{sec:4}

Over the course of the SDSS-IV survey, periodic validation rounds took place as new data became available.  Upon assembly of a new ``run'', a call for inspectors was announced to the entire collaboration.  A minimum number of groups consisting of 50 targets each were assigned to insure that at least 2 votes are recorded in every case, with no restriction on the maximum number of groups an inspector wants to validate.  Below is the summary for each sample making up the final catalogue.  The name of the run signifies the cut off date for which data was included.  General details for each run are found in Table~\ref{tab:validation_results}. 

\subsection{Run 2016-07-04}

Just for the purpose of breaking up the work load into more equal parts, each round was defined by a richness cut.  Round one consisted of all observed candidates, marked as complete, with optical richness ($\mathrm{\lambda}_\text{OPT}$) greater than 30. Candidates marked as complete are systems with as many observations as their allocated amount of spectroscopic fibres. Except for marginal changes due to reprocessing of the data, their observational status is therefore definitive in the context of the project. A total of 573 candidates were inspected in this round.  This round makes up the entirety of the DR14 catalogue release \citep{dr14}.  For the second round, only completed candidates with the cut 20 $<$ $\mathrm{\lambda}_\text{OPT}$ $<$ 30 were considered.  This amounted to 344 new inspections.  For the final round, the remaining 618 candidates with 10 $<$ $\mathrm{\lambda}_\text{OPT}$ $<$ 20 were inspected.  The two available inspectors for this round disagreed on a total of 21 candidates.  Due to there being no way to break the disagreements, two additional inspectors were added during the reconciliation round to find a consensus for these candidates.  A final total of 1052 validated components were found.  The full breakdown for each round is presented in Table~\ref{tab:validation_results}.

\subsection{Run 2018-04-27}

This run consists of the next 22 months of observations, all were new to the inspectors.  The first three rounds were broken up into the same richness bins as the previous run.  Round 1 consisted of inspections of 434 candidates, round 2 consisted of inspections of 216 candidates, and round 3 consisted of inspections of 432 candidates.   The 4th round was initiated as a special round to catch up on missing candidates.  It was determined that candidates labeled incomplete in the original SEQUELS pilot area \citep{cle16} no longer had the potential to be re-observed due to changing survey strategy and targeting.  Every candidate in this area was included for inspection, totaling 193.  Only two inspectors were available for this round again.  A third inspector voted to break the disagreement for the 4 candidates with no consensus.  A final total of 1001 validated components were found.  Table~\ref{tab:validation_results} contains more details for each round within this run.

\subsection{Run 2018-12-04}

A third run was initiated in order to reassess the effect of major changes to the spectra processing pipeline \citep{ahu20}.  Included in this run are all completed candidates observed over the previous 7 months and all candidates from the previous two runs where 1 or more galaxies show a different redshift, redshift error, or warning flag.  Round 1 consisted of a total of 383 candidates. Among them, 200 were new inspections and 183 were reanalyses of previous systems. Among clusters validated in the previous runs, 52 were validated again and one was unvalidated thanks to one galaxy redshift change in its red-sequence and a more careful examination. Among the clusters unvalidated in previous run, 88 became validated.  A final total of 299 validated components were found.  See Table~\ref{tab:validation_results} for more details.  

\subsection{Run 2019-03-22}

The final run took place after the completion of the SPIDERS survey.  Round 1 consisted of the remaining 818 candidates with completed observations.  Round 2 consisted of the remaining 140 candidates that had partially completed observations.  About 720 were new inspections and little less than 250 were reanalyses of systems inspected in earlier runs. Among the previously validated systems, 98 were validated again, none were unvalidated. Among the previously unvalidated systems, 72 became validated. A final total of 717 validated components were found.  An additional third round was needed during the construction of the final catalogue for re-inspecting multi-component clusters that did not fall into any of our classification categories.  The details of this can be found in section~\ref{sec:5}.  Additional details of the rounds are presented in Table~\ref{tab:validation_results}.

\subsection{Inspection statistics}

The mean number of inspectors per cluster candidate in this sample was approximately 3.  Over the full sample of 2740 validated clusters, there are only 71 instances where a full consensus was not reached between inspectors.  For spectroscopic redshift determination, there are 622 instances where there was not exact agreement.  The mean spread for these instances is 0.00049 (147 km s$^{-1}$), with the maximum disagreement being 0.0055 (1\,650 km s$^{-1}$).

Comparing to the automatic membership assignment, there are 955 instances where the redshift was changed after inspection.  Within this sample, there are 228 cases where inspectors additionally broke the candidates up into 2, 3, or 4 separate components.

\begin{figure}
    \centering
    \includegraphics[width=\columnwidth]{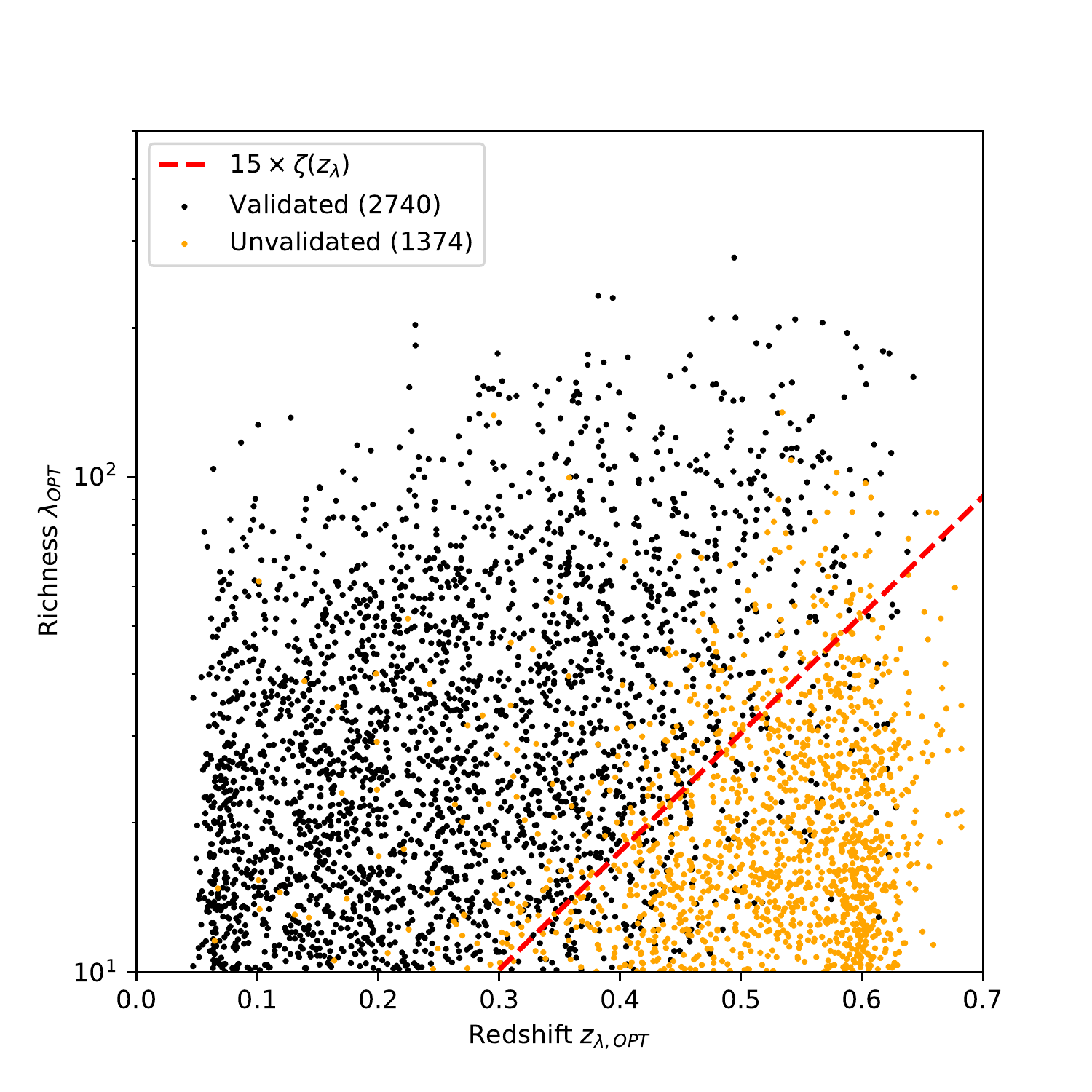}
    \caption{The validation results of all inspection rounds are plotted showing richness versus photometric redshift.  Validated clusters are shown in black, unvalidated candidates are shown in yellow.  Only the highest richness (mass) clusters are validated with increasing redshift. The dashed red line roughly indicates the transition between mostly validated and mostly unvalidated, where $\zeta(z_{\rm{\lambda}})$ is the redshift dependent photometric depth correction factor for richness estimates.}
    \label{fig:validation_fraction}
\end{figure}

The chances for a candidate to be validated were highly dependent on redshift, richness, and number of spectra obtained.  Figure~\ref{fig:validation_fraction} shows the optical richness versus photometric redshift for the all validated and unvalidated clusters from the entire program.  There is very little dependence below a redshift of 0.3.  Out to a redshift of 0.4 though, most clusters below richness 20 are unvalidated.  This trend increases until a redshift of 0.6 where almost all clusters are unvalidated except the most extreme objects ($\mathrm{\lambda}_\text{OPT}$ $>$ 100).  This follows approximately $15 \times \zeta(z_{\rm{\lambda}})$, {where \bf $\zeta=e^{5.5(z-0.35)}-0.12$ (at $z>0.37$)} is the redshift-dependent scaling adopted by redMaPPer to account for galaxies brighter than the $0.2L^*$ limit, but are fainter than can be detected in SDSS \citep[][their Fig.~19]{ryk14}. This empirical delineation is materialized with the red dashed line in Fig.~\ref{fig:validation_fraction}; it reflects that about fifteen $L>0.2L^*$ galaxies must appear in SDSS images in order for a cluster to be validated. The red line has been obtained by setting a requirement on targeting at least 8 of the cluster member candidates, as with an average success rate of measuring the redshift in accordance with cluster redshift of 0.7, which leads to at least 5 confirmed member galaxies.

\begin{figure}
    \centering
    \includegraphics[width=\columnwidth]{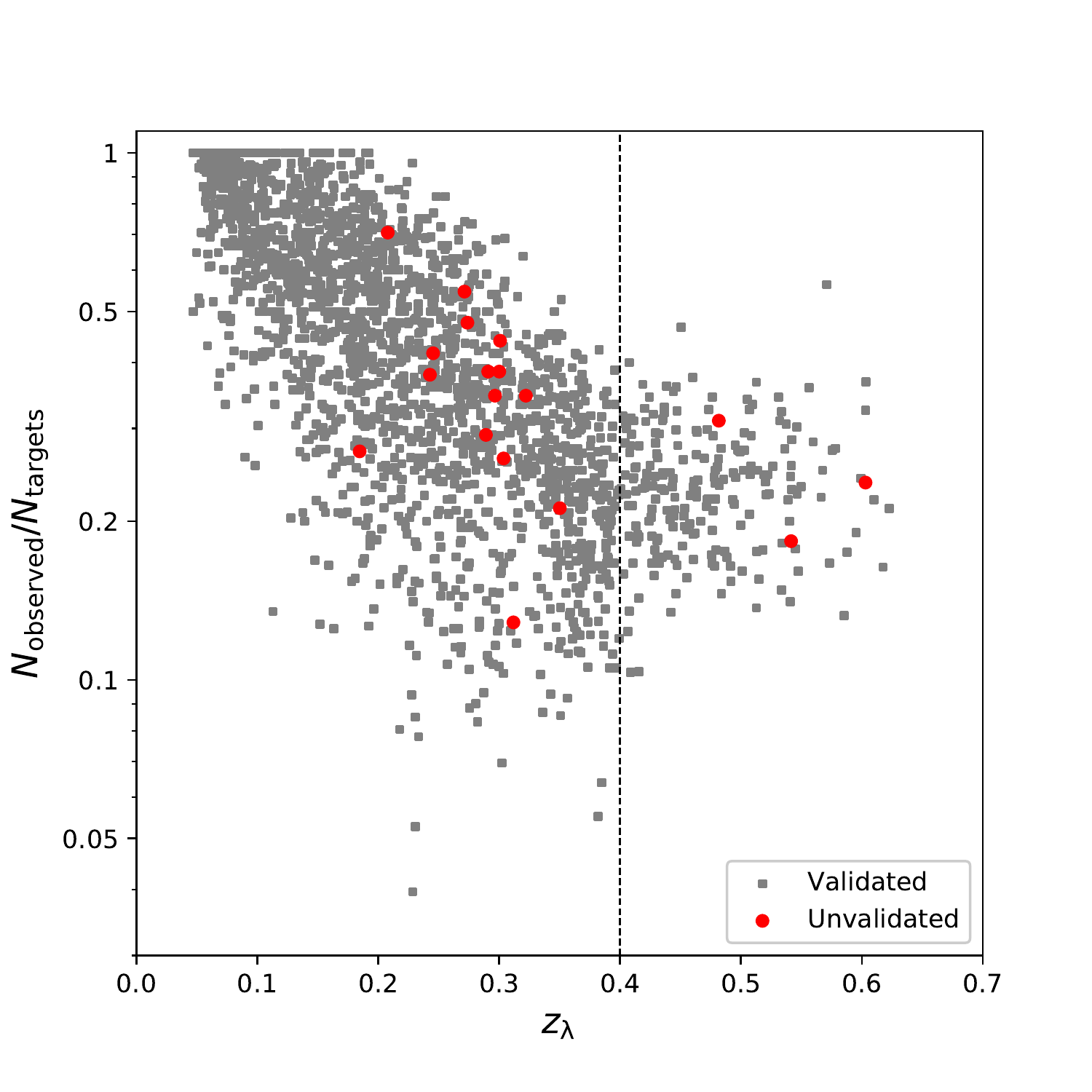}
    \caption{The ratio of observed members (i.e., spectroscopic redshift was obtained) to total red-sequence members for each cluster candidate (where $N_{\rm{observed}} \geq 9$) is plotted against the photometric redshift.  Validated points are shown in grey and unvalidated points are shown in red.  Over this entire sub-sample, less than 1\% of candidates are unvalidated.  At a redshift of 0.4 (black dashed line), the maximum observing efficiency appears to flatten out at an approximate 1:3 ratio.}
    \label{fig:validation_nhasz}
\end{figure}

Figure~\ref{fig:validation_nhasz} illustrates the targeting efficiency.  We present the ratio of observed galaxies to all red-sequence member candidates versus the cluster photometric redshift, where a minimum of 9 spectra were available per cluster.  Below a redshift of 0.2, we are able to reach a maximum 1:1 ratio.  This maximum ratio drops off to 1:2 out to redshift 0.3, and further to 1:3 out to redshift 0.4.  Beyond that, no trend is discernible.  The red points represent all clusters that were unvalidated with a minimum of 9 spectra obtained.  We achieved a $>$99\% validation rate across the entire sample at this level.  Considering only clusters at $z<0.3$, the validation rate improves to $~$99\% when at least 5 spectra are obtained.

\section{Catalogue Construction} \label{sec:5}

Basic X-ray properties for every cluster that has been validated have been calculated.  We report the fluxes corrected for Galactic absorption. In performing this calculation, we assume a constant spectral shape of the source, and perform a correction for this assumption as a part of the K-correction, which accounts for the effect of the source spectral shape (defined by temperature of the emission and the redshift). We use the XXL $L-T$ \citep{giles} relation and $M-T$ \citep{lieu} relation and a flat $\Omega_m=0.3$ $\Lambda$CDM cosmology in the calculations of $L_X$, as those agree with the CODEX weak lensing work of \citet{kettula15}.

The fluxes are measured within the apertures defined to contain a significant flux measurement, with respect to the background, but constrained to be within $12^\prime$. As these do not necessarily cover the whole cluster, we extrapolate the measurement using the expected surface brightness profiles of clusters. The procedure is described in \cite{finoguenov07}  and has to be applied iteratively, allowing the cluster size to change with iterations, based on the obtained $L_X$. For nearby clusters with large corrections factors (greater than 2), we re-extracted fluxes using much larger apertures, extending to $48^\prime$, confirming the results.

Further filtering must be carried out before merging the separate validations runs into the final catalogue.  The first step is to classify all clusters that have been validated with multiple components in the line of sight.  Inspectors are tasked with ordering the components from most to least significant.  The most significant component is considered the main gravitating halo along the line of sight, and this is defined as the component with the most spectroscopic members associated with it.  For instances where membership is equal between components, the component that is located at a lower redshift is considered the most significant.  Inspectors may also choose to ``re-merge'' components if they believe the cluster is experiencing a near 1:1 merger, as there is no clear indication of which part is the most significant.  This results mainly when components are within 4000 km/s of each other.  Round 3 of run 2019-03-22 was the re-inspection of clusters deemed to be likely experiencing this type of merger.  For these cases, the redshift of gravitating halo is accurate, but the velocity dispersion reported is likely unreliable as a mass proxy.  Once the classification is complete, agreement is required by all inspectors before a cluster is entered into the final catalogue.  Where disagreements exist still, further discussion takes place until a unanimous decision is reached.  Only the main component is entered into the catalogue.  A flag in the final catalogue, $\rm{NCOMPONENT}$, indicates the number of visually identified components, or $-1$ for clusters identified as mergers.

The next step is to concatenate the separate catalogues while simultaneously removing duplicate clusters.  Due to some clusters being inspected multiple times throughout the validation process, we took care to choose the newest evaluation as the one to enter our catalogue.  Following this, duplicates with unique IDs are also filtered out.  As the CODEX catalogue is compiled from two separate X-ray detection and RedMaPPer runs, a few rare instances occur where candidates are entered twice into the target list under different IDs.  These are determined by matching clusters based on RA, Dec, and spectroscopic redshift.  The few cases are checked visually as to whether they are sharing the same member galaxies.  These are removed, keeping the ID from the second RedMapper run (IDs starting with 2).

\begin{figure}
    \centering
    \includegraphics[width=\columnwidth]{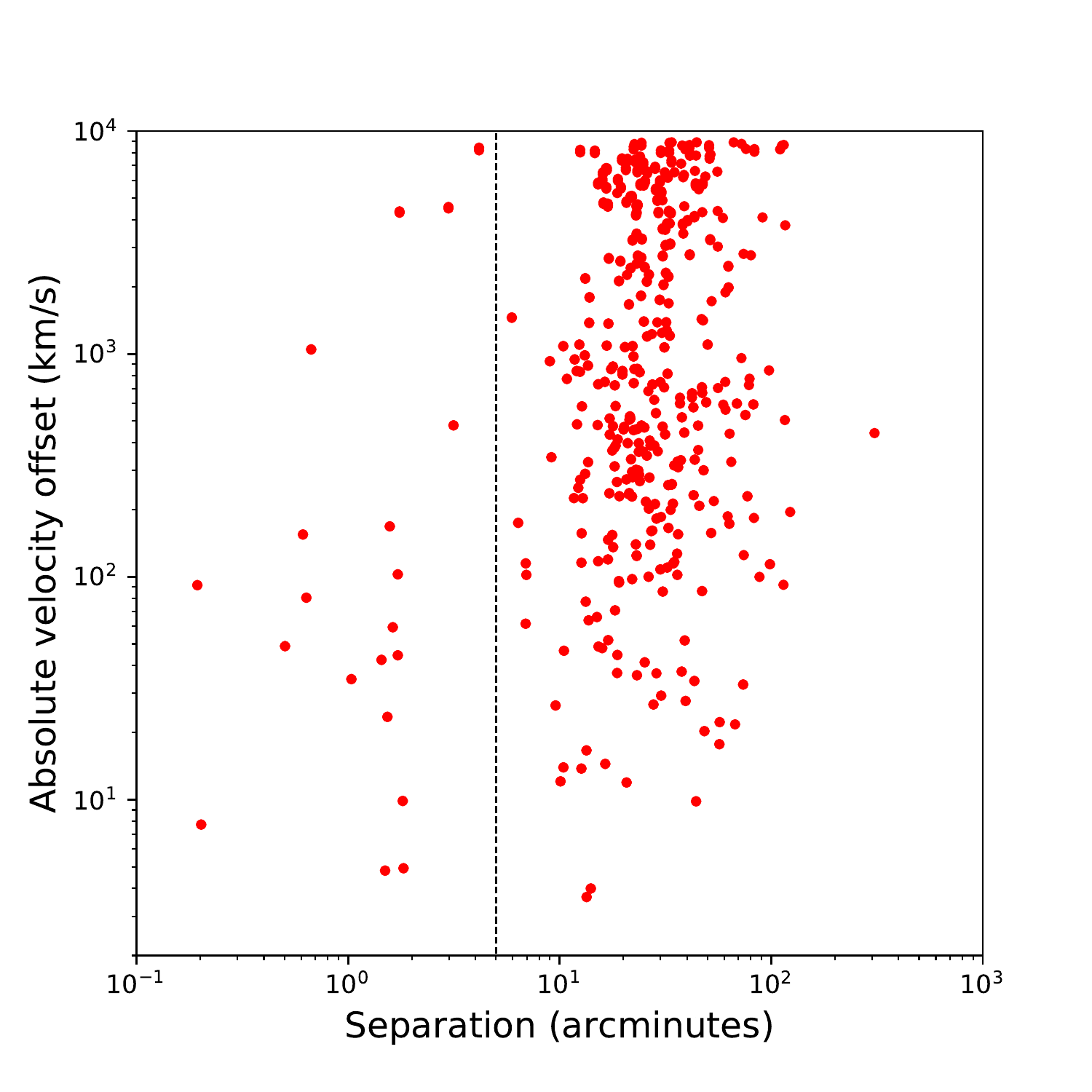}
    \caption{Absolute velocity offset versus projected sky separation between cluster pairs before cleaning the sample.  The pairs to the left of the vertical dashed line were visually inspected individually for common galaxy members. Some clusters are represented multiple times due to proximity of more than one other cluster.}
    \label{fig:nearby}
\end{figure}

Next, we attempted to remove all clusters still remaining that were validated under unique IDs, but are still likely to be the same structure being counted twice.  Clusters are matched with any other cluster in the sample that is within 5 times R$_{200}$ and $0.03(1+z)$ in redshift ($<9000$ km s$^{-1}$) of itself.  In Figure~\ref{fig:nearby}, we plot the absolute velocity offset against the projected sky separation for each pair of clusters.  It is likely that some these clusters are in some phase of merging.  Most pairs are found to be separated within 10 to 100 arcminutes, though a small number are within a few arcminutes of each other.  With centres being based on RASS detections, it is not possible to distinguish the difference between X-ray sources with less than 2 arc minute separation.  Upon visually inspecting pairs within 5 arcminutes of each other, we discover most pairs have inter-mixed galaxy members.  In these cases, we enter the pair into the final catalogue as a single object.

For the final step, we make one last check for mistakenly unvalidated candidates.  Candidates that appear to be outliers from the rest of the sample (see the red points in Figure~\ref{fig:validation_nhasz}) are visually inspected one last time.  We collaboratively agree that the data available is not sufficient enough to determine a cluster redshift.  This finalises the catalogue.

\section{The Catalogue} \label{sec:6}

The full SPIDERS cluster catalogue of optical properties, redshifts, dynamical properties, and X-ray properties can be found online\footnote{\texttt{https://www.sdss.org/dr16/data\_access/value-added-catalogs/?vac\_id=spiders-x-ray-galaxy-cluster-catalogue-for-dr16}}.  In total, our catalogue contains 2740 visually inspected galaxy clusters.  Contained within these clusters are 33,340 individual galaxy members.  Along with this cluster catalogue, we have released the database of redshifts used for identification \citep{ahu20}, the targeting scheme \citep{cle16}, and the full target catalogue\footnote{\texttt{https://www.sdss.org/dr16/data\_access/value-added-catalogs/?vac\_id=spiders-target-selection-catalogues}}.

\begin{table*}
    \centering
    \begin{tabular*}{\textwidth}{l@{\extracolsep{\fill}}llr}
        \hline
        \multicolumn{1}{l}{Column} & \multicolumn{1}{l}{Unit} & \multicolumn{1}{c}{Description} & \multicolumn{1}{r}{Example} \\
        \hline
        \texttt{CLUS\_ID} (1) & & SPIDERS/CODEX identification number & $1\_6939$ \\
        \texttt{NCOMPONENT} (2) & & Number of validated components, or merger flag & 1 \\
        \texttt{CODEX} (3) & & The CODEX cluster candidate unique identifier & 26240 \\
        \texttt{RA} (4) & deg & CODEX X-ray detection right ascension (J2000) & 233.106 \\
        \texttt{DEC} (5) & deg & CODEX X-ray detection declination (J2000) & 39.003 \\
        \texttt{RA\_OPT} (6) & deg & CODEX optical detection right ascension (J2000) & 233.083 \\
        \texttt{DEC\_OPT} (7) & deg & CODEX optical detection declination (J2000) & 39.024 \\
        \texttt{LAMBDA\_CHISQ\_OPT} (8) & & Richness $\mathrm{\lambda}_\text{OPT}$ of the CODEX optical detection & 38.8 \\
        \texttt{Z\_LAMBDA} (9) & & Photometric redshift (z$_\lambda$) of the CODEX optical detection & 0.27 \\
        \texttt{Z\_LAMBDA\_ERR} (10) & & Uncertainty on z$_\lambda$ & 0.01 \\
        \texttt{NMEM} (11) & & Number of objects in the CODEX red-sequence & 50 \\
        \texttt{NOKZ} (12) & & Number of red-sequence members with a spectroscopic redshift & 14 \\
        \texttt{SCREEN\_CLUZSPEC} (13) & & Galaxy cluster redshift, assigned after visual inspection & 0.2863 \\
        \texttt{SCREEN\_CLUZSPEC\_ERR} (14) & & Bootstrap uncertainty on \texttt{SCREEN\_CLUZSPEC} & 0.0012 \\
        \texttt{SCREEN\_CLUZSPEC\_SPREAD} (15) & & Dispersion in the inspection cluster redshifts & 0 \\
        \texttt{SCREEN\_CLUZSPEC\_GAP} (16) & km s$^{-1}$ & Gapper estimate of the cluster velocity dispersion & 790.0 \\
        \texttt{SCREEN\_CLUVDISP\_BWT} (17) & km s$^{-1}$ & Square root of the bi-weight variance velocity dispersion & 805.5 \\
        \texttt{SCREEN\_CLUVDISP\_BEST} (18) & km s$^{-1}$ & Value of the ``best velocity dispersion'' & 790.0 \\
        \texttt{SCREEN\_NMEMBERS\_W} (19) & & Weighted number of red-sequence members identified as members & 11 \\
        \texttt{STATUS} (20) & & Validation status of the cluster assigned by the visual inspector & validated \\
        \texttt{NINSPECTORS} (21) & & Number of individual inspections for this system & 3 \\
        \texttt{NVALID} (22) & & Number of inspectors validating this system as a galaxy cluster & 3 \\
        \texttt{LX0124} (23) & ergs s$^{-1}$ & Luminosity in the (0.1-2.4) keV band of the cluster, aperture $R_\text{500c}$ & $1.5\times10^{44}$ \\
        \texttt{ELX} (24) & ergs s$^{-1}$ & Uncertainty on \texttt{LX0124} & $0.5\times10^{44}$ \\
        \texttt{R200C\_DEG} (25) & deg & Apparent $R_\text{200c}$ radius of the galaxy cluster & 0.088 \\
        \texttt{FLUX052} (26) & ergs s$^{-1}$ cm$^{-2}$ & Galaxy cluster X-ray flux in the 0.5-2.0 keV band & $3.9\times10^{-13}$ \\
        \texttt{EFLUX052} (27) & ergs s$^{-1}$ cm$^{-2}$ & Uncertainty on \texttt{FLUX052} & $1.2\times10^{-13}$ \\
        \texttt{MCXC} (28) & & Identifier in the MCXC catalogue \citep{pif11}, if present & n/a \\
        \texttt{ANAME} (29) & & Alternative name in \citep{pif11}, if present & n/a \\
        \hline
    \end{tabular*}
    \caption{Descriptions of the columns for the validated SPIDERS/CODEX galaxy cluster catalogue.  The full VAC, and all others for DR16, are available online at: \texttt{https://www.sdss.org/dr16/data\_access/value-added-catalogs/}}
    \label{tab:catalogue_columns}
\end{table*}

The names of the included columns are presented in Table~\ref{tab:catalogue_columns} with units, a short description, and an example of the data formatting.  The columns of the catalogue provide: (1) the SPIDERS name, (2) the total number of individually identified cluster components along the line-of-sight, or a value of $-1$ is given if the multiple components are undergoing a major merger, agreed upon by at least two inspectors, (3) the CODEX target candidate name, (4) and (5) the right ascension and declination for the epoch J2000 in degrees of the RASS X-ray detection, (6) and (7) the right ascension and declination for the epoch J2000 in degrees of the optical centre determined by redMaPPer, (8) the redMaPPer richness based on aperture at the optical centre, (9) and (10) the photometric redshift of the red-sequence and the associated error estimation, (11) the total number of member candidates in the CODEX targeting catalogue, (12) the total number of member candidates with a spectroscopic redshift, (13) the assigned redshift for the galaxy cluster after visual inspection, (14) the bootstrap uncertainty on the assigned galaxy cluster redshift, (15) the dispersion in redshift between inspectors as described in Section 3, (16) the velocity dispersion calculated using the gapper estimate (GAP) in units of km s$^{-1}$, (17) the velocity dispersion using the square root of the bi-weight variance (BWT) in units of km s$^{-1}$, (18) the best velocity dispersion in units of km s$^{-1}$ based on \citet{bee90} where we report GAP if less than 15 members are used in the calculation and BWT for well sampled systems of 15 or more members, (19) the sum total of weighted membership flags of red-sequence galaxies determined after inspection, (20) the validation status assigned after inspection reconciliation, (21) the number of individual inspections for the given validated galaxy cluster, (22) the number of inspectors to give a status of validated, (23) and (24) the calculated X-ray luminosity in units of ergs s$^{-1}$ for the 0.1-2.4 keV band using an aperture radius of $R_\text{500c}$ and the associated uncertainties as described in Section 5, (25) the $R_\text{200c}$ radius of the galaxy cluster in units of degrees, (26) and (27) the measured X-ray flux in units of ergs s$^{-1}$ cm$^{-2}$ for the 0.5-2.0 keV band and associated uncertainties as described in Section 5, (28) and (29) the matched identifier in the MCXC catalogue or an alternate name given by \citet{pif11} in cases where the galaxy cluster appears in both catalogues.

The optical RA and DEC coordinates given in columns (6) and (7) are determined by running redMaPPer a second time with a relaxed constraint on the centre.  The uncertainties on the initial X-ray position of RASS sources is approximately 3 arcmin, which the red-sequence finder is allowed to vary within in order to optimize the centre.  The richness given in column (8) refers to the redMaPPer richness estimator value determined, prior to SPIDERS follow-up, from red-sequence members found using the optical centre.  It is equal to the sum of the membership probabilities over the field of view, the details of which are described in \citet{ryk14}.  There are additional factors taking into account the unobserved, masked areas within the search aperture and the photometric depth correction, modeled as an exponential increase towards high-$z$ \citep{fin20}.  The radii $R_\text{500c}$ and $R_\text{200c}$ referenced in columns (23), (24), and (25) are determined using the $L_\text{X}$-$M$ scaling relations of \citet{lea10}.

\begin{figure}
    \centering
    \includegraphics[width=\columnwidth]{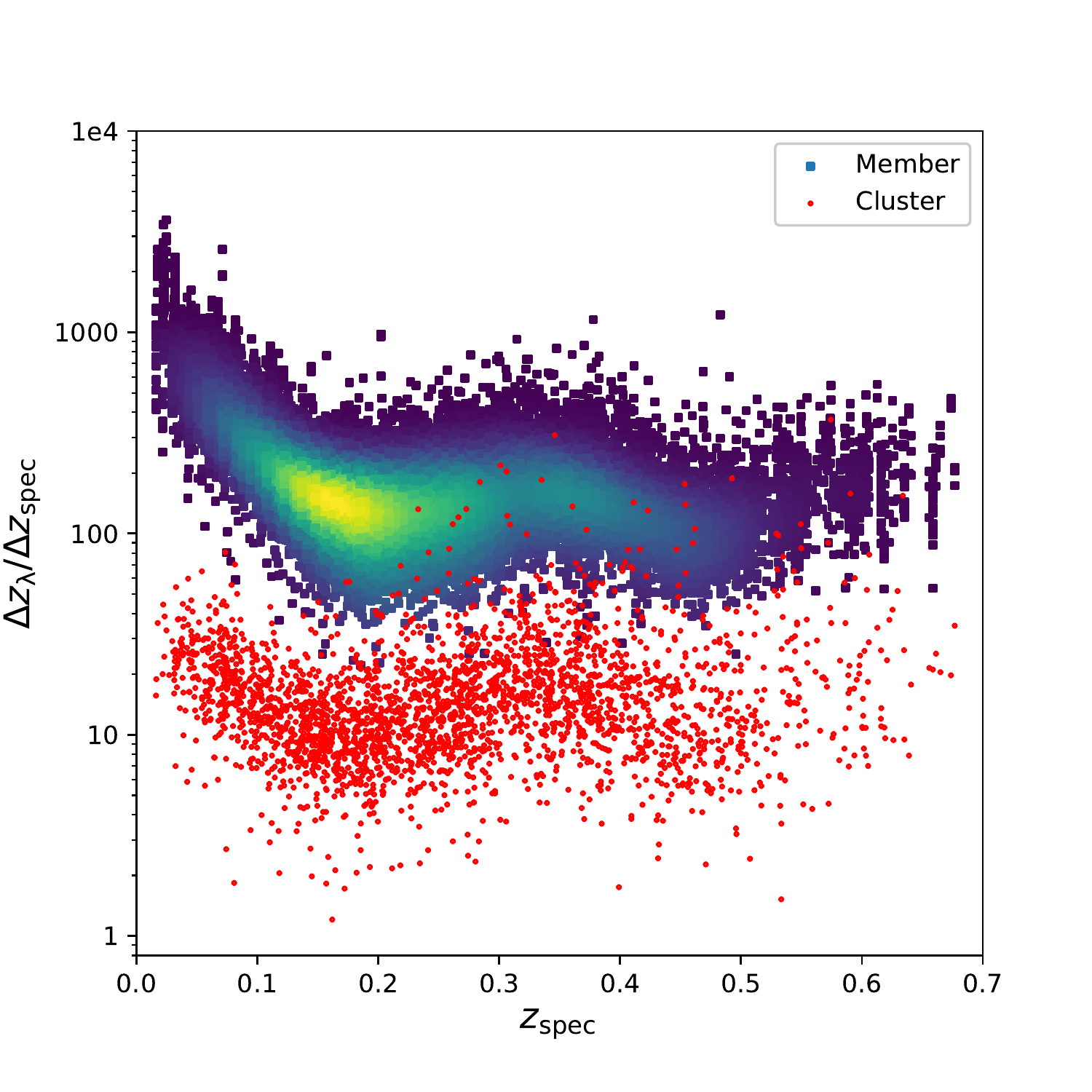}
    \caption{Ratio of photometric to spectroscopic redshift errors plotted against the spectroscopic redshift.  Clusters are represented by red circles.  The individual member galaxies are represented as squares highlighted by point density.}
    \label{fig:redshift_err}
\end{figure}

Here we present a few examples of the improvements this catalogue achieves due to large scale spectroscopic follow-up.  Figure~\ref{fig:redshift_err} highlights the improved accuracy on redshifts.  We plot the ratio of photometric to spectroscopic redshift errors against the spectroscopic redshift.  The typical statistical uncertainty on cluster redshifts is $\Delta z /(1+z) = 6 \times 10^{-4}$ (Clerc et al. accepted).  We demonstrate that this translates into an improvement of a factor of $\sim10$ compared to the photometric estimation.  Redshifts below 0.1 are improved even greater due to the known short comings of red-sequence finders on nearby objects \citep{ryk14}.  Overall, this is achieved due to a redshift improvement by a factor of $\sim100$ for individual galaxies, demonstrated by density shaded points.

\begin{figure}
    \centering
    \includegraphics[width=\columnwidth]{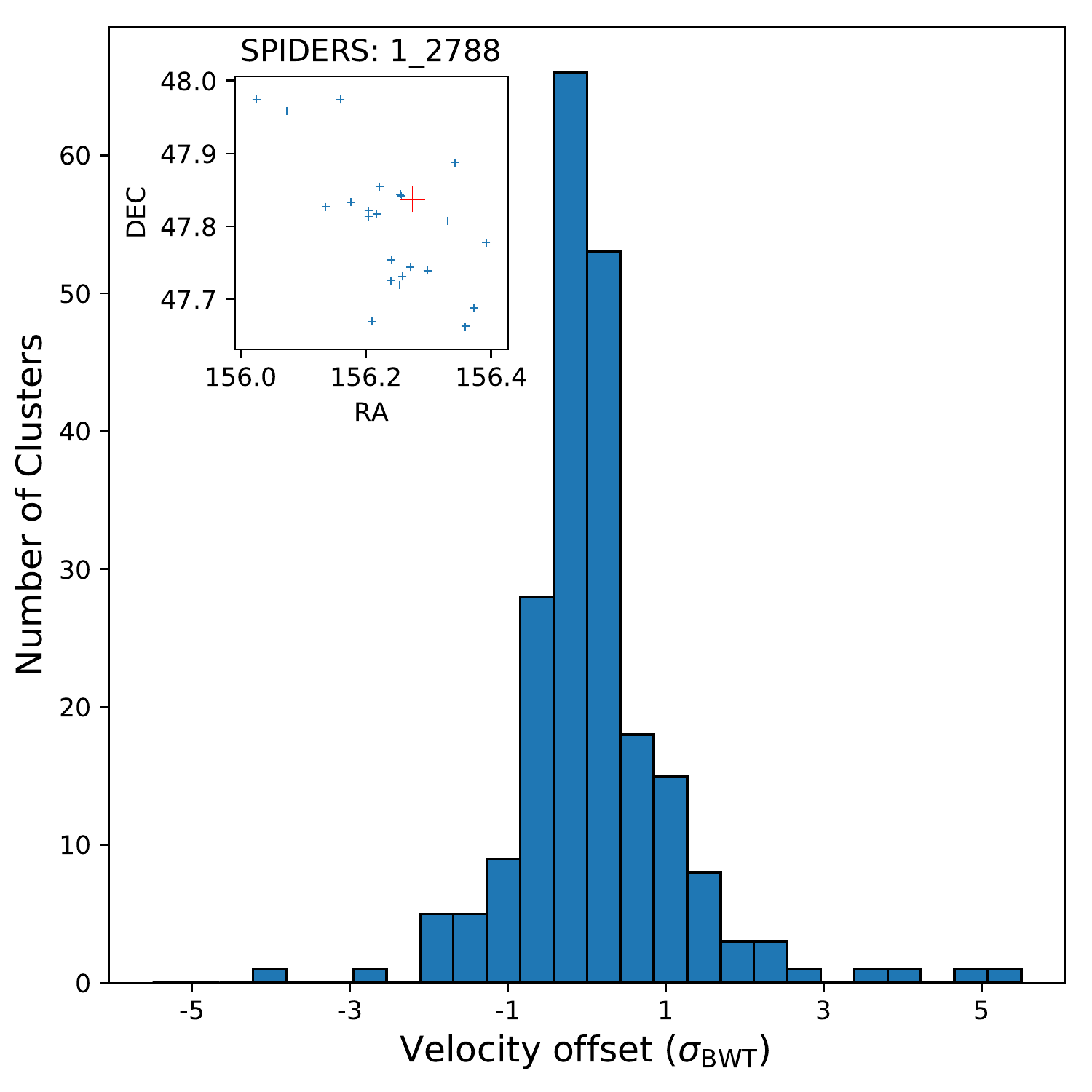}
    \caption{Distribution of velocity offset between matched SPIDERS and MCXC cluster, normalised by $\sigma_{\rm{BWT}}$.  Offset is determined by the same calculation of proper velocity used during visual inspection.  Inset: Projected sky coordinates of SPIDERS member galaxies (blue crosses) and the X-ray peak centre from the MCXC catalogue (red cross).  The example matched cluster (\texttt{1\_2788/J1025.0+4750)} has a discrepancy in proper velocity of approximately 5$\sigma$.}
    \label{fig:zdiff}
\end{figure}

When comparing our redshift determinations to the 220 clusters in common with the MCXC catalogue \citep{pif11}, we find good agreement in general with only a handful of outliers.  Figure~\ref{fig:zdiff} illustrates the distribution of the velocity offset between SPIDERS redshifts and MCXC redshifts, normalized by $\sigma_{\rm{BWT}}$.  Only 5 systems in the sample are offset by more than 3$\sigma$, a minimum requirement that suggests these matches would not be considered one object during visual inspection.  On re-inspection of these outliers, all but one case are of systems that are highly sampled (the other system only having 5 redshifts available), and only one system has evidence for multiple components along the line-of-sight.  Inset into Figure~\ref{fig:zdiff}, we have plotted an example of a cluster match with an approximate 5$\sigma$ offset.  The red cross indicates the MCXC centre (X-ray peak) of the cluster, where the blue crosses are the individual member galaxies, 23 total, used in the calculation of the cluster redshift.  No projected offset is apparent, indicating both catalogs are referring to the same extended X-ray source.  Without knowing the exact method for determining the individual redshifts in MCXC, we cannot explain the discrepancy in velocity offset, though we can conclude our measurement is robust.  A detailed comparison of our redMaPPer sample with Planck clusters \citep{roz14,roz15} has also been carried out.  These results are encouraging and a thorough comparison of the spectroscopic properties of all available overlapping surveys is required, but is beyond the scope of this current work.

\begin{figure}
    \centering
    \includegraphics[width=\columnwidth]{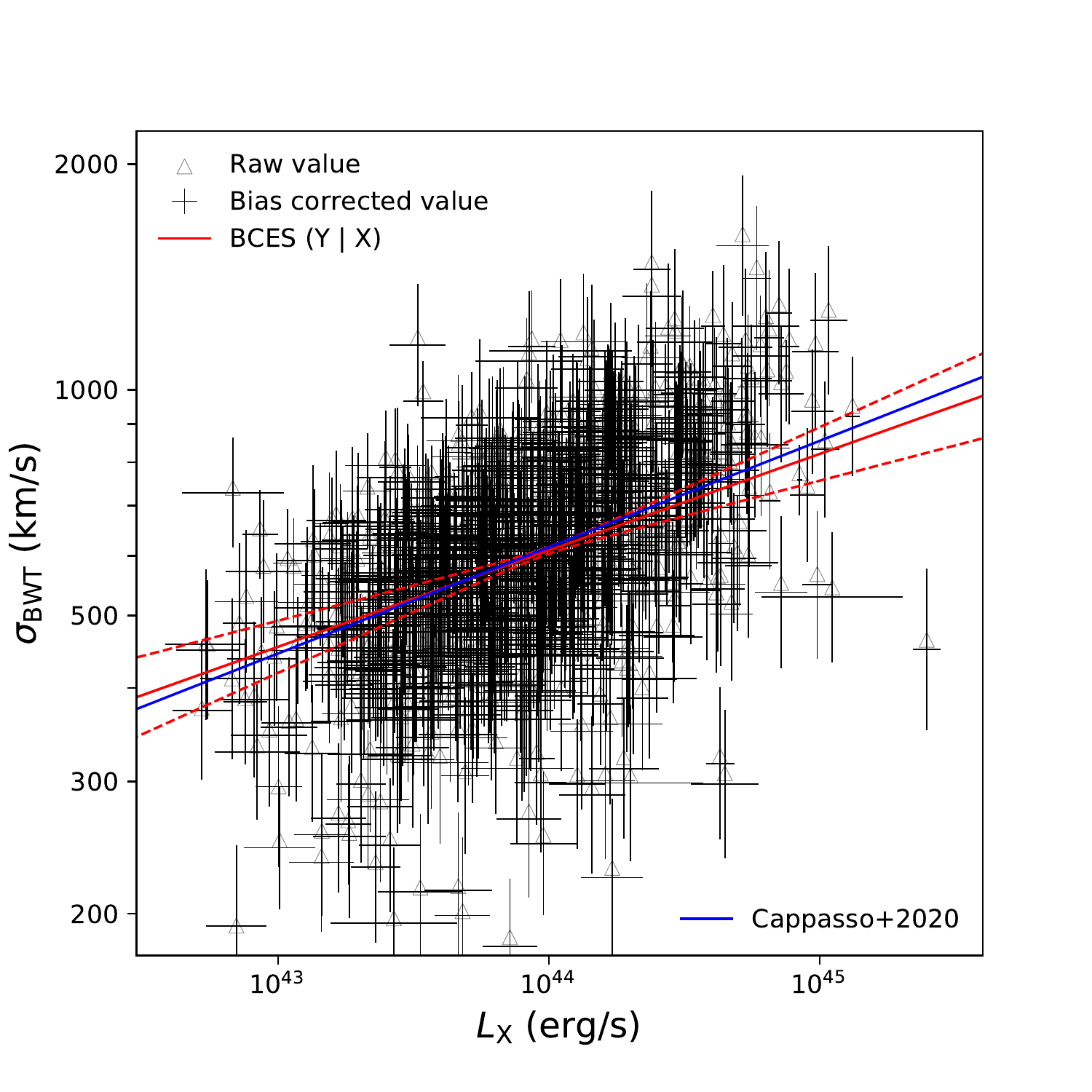}
    \caption{$\sigma - L_{\rm{X}}$ relation for SPIDERS clusters.  The plotted points are clusters with 15 or more validated members for $z < 0.4$.  The triangles represent the raw bi-weight variance calculations.  The crosses represent the bias corrected values together with their uncertainty.  The solid and dashed red lines show the BCES fit to the bias corrected values and the 1$\sigma$ uncertainty range.  The solid blue line represents the translated $L_{\rm{X}} - M_{\rm{200c}}$ scaling relation from \citet{cap20}.}
    \label{fig:sig_lx}
\end{figure}

In Figure~\ref{fig:sig_lx}, we present the updated $\sigma - L_{\rm{X}}$ relation as presented in \citet{cle16}.  This improves the previous DR14 release by increasing the highly sampled systems from 39 clusters to 755 clusters.  We define a highly sampled system as having 15 or more spectroscopic members used in the calculation of velocity dispersion as well as only one detected component along the line of sight to insure no external force is affecting the measurement.  We also make a cut in redshift at $z<0.4$, as the sample above that redshift (32 clusters only) is highly incomplete.  The raw values used in the $\sigma - L_{\rm{X}}$ plane are the reported values in column (17) of the catalogue.  The bias corrected values are calculated using an updated approach from the previous release \citep{cle16}.  Details of this are described in section~\ref{sec:7}.  Associated errors are calculated using the method of re-sampling observations of clusters with a well determined velocity dispersion using a high number of spectroscopic observations \citep{rue14}.  We again compute the best-fitting power law to the bias corrected data points using the BCES method \citep{akr96}, with the $y$-axis as the dependent variable, to illustrate our samples trend.  We fit constants A and B defined as,
\begin{equation}
    \log \left(\frac{\sigma_{\rm{BWT}}}{700~\rm{km~s^{-1}}}\right) = A + B  \log \left(\frac{L_{\rm{X}}}{10^{44} \rm{ergs~s^{-1}}}\right),
\label{eq:sig_lx}
\end{equation}
where we find the best fits to be $(-5.91 \pm 1.47) \times 10^{-2}$ and $(1.29 \pm 0.35) \times 10^{-1}$, respectively.  We find a standard deviation of 0.13 along the vertical axis of the fit.  Over-plotted is the comparison to the $L_{\rm{X}}-M_{\rm{200c}}-z$ relation of \citet{cap20}.  Both samples are drawn from the same catalogue, though they use an earlier release with no redshift cut, for a total of 344 systems.  Also used are an updated measurement of X-ray luminosity \citet{fin20}, where flux is extracted through larger apertures for nearby clusters instead of the smaller fixed aperture of our catalogue.  We evaluate the relation at fixed mass and redshift, $M_{\rm{200c}} = 3 \times 10^{14}$ and $z = 0.16$, the approximate mean of both samples.  Following \citet{Carlberg1997aa}, we scale with velocity dispersion as,
\begin{equation} \label{eq:sigma-mass}
    \sigma = \frac{10 R_{\rm{200}}  H(z)}{\sqrt{3}},
\end{equation}
\begin{equation}
    R_{\rm{200}} = \left( \frac{3 M_{\rm{200c}}}{4\pi  200\rho_{\rm{cr}}} \left(\frac{H(z)}{100}\right)^{-2} \right)^{1/3}.
\label{eqn:sig_rich}
\end{equation}
With twice the number of systems with a higher quality selection, it is encouraging that out simple analysis shows good agreement.  There is also no effect on the fit whether using X-ray luminosity from our catalogue, or the updated values.  A proper derivation utilising a fully consistent statistical treatment is the important next step.

\begin{figure}
    \centering
    \includegraphics[width=\columnwidth]{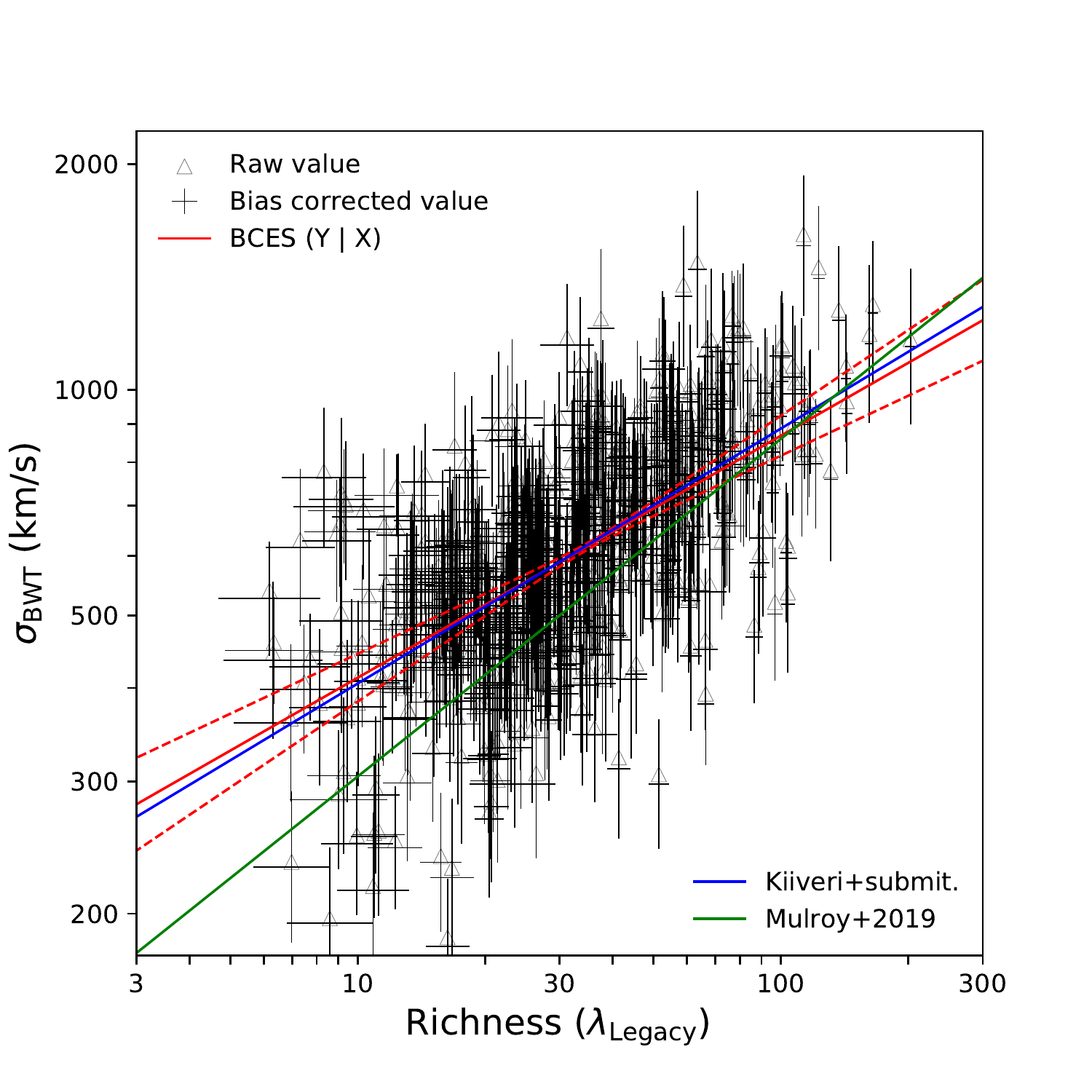}
    \caption{$\sigma$-$\lambda$ relation for SPIDERS clusters.  This sample is drawn the same cluster sample used in Figure~\ref{fig:sig_lx}.  Raw and bias corrected bi-weight variances are represented by triangles and crosses, respectively. The solid and dashed red lines shows the BCES fit and $1\sigma$ uncertainty for the bias corrected values and updated richness calculations.  The solid blue and green lines represent the translated predictions from \citet{kii20} and \citet{mul19}.}
    \label{fig:sigma-lambda}
\end{figure}

We also obtain a fit for the $\sigma - \lambda$ relation, presented in Figure~\ref{fig:sigma-lambda}.  Starting with the same sample, we find all clusters overlap with the DESI Legacy imaging surveys \citep{dey19}, for a total of 611 clusters.  With deeper imaging data, a more accurate calibration of richness is achieved.  RedMaPPer is re-run at the position of the optical centre for all CODEX clusters, allowing for a new centre to be found \citep{IderChitham2020}.  This improvement reduces the up-scatter of low mass systems that occurs due to shallow imaging, especially at high redshift.  At the mean redshift of our sample, this improvement scales as $\lambda_{\rm{Legacy}} \sim 0.89 \cdot \lambda_{\rm{SDSS}}$.

We aim to compute the best-fitting power law to these new data using the same BCES method as before to illustrate the sample trend, compare to predictions from the literature and look for evidence of redshift evolution.  We fit constants A and B defined as,
\begin{equation}
    \log \left(\frac{\sigma_{\rm{BWT}}}{700~\rm{km~s^{-1}}}\right) = A + B  \log \left(\frac{\lambda_\text{Legacy}}{35}\right),
\end{equation}
where we find the best fits to be ($-5.36 \pm 0.45) \times 10^{-2}$ and ($3.23 \pm 0.57) \times 10^{-1}$, respectively.  We find an overall standard deviation of 0.11 for the fit.  Dividing the sample into low (< 35) and high (> 35) richness, the standard deviation along the vertical axis is 0.11 and 0.12, respectively, indicating it is relatively constant along the fit.  For comparison, we translate results based on studies of mass and richness for different cluster studies.  \citet{kii20} studied a small sample of high-$z$, high-$\lambda$ clusters using deep CFHT multi-band imaging.  They use a hierarchical Bayesian model to determine their $\lambda_{\rm{CFHT}} - M_{\rm{WL}}$ relation.  We have compared the legacy richness measurements to the their own CFHT measurements, finding the richness scaling as $\lambda_{\rm{Legacy}} \sim 0.94 \cdot \lambda_{\rm{CFHT}}$.  Using this scaling, their fit compares favorably with our low-z sample.  The good agreement, even with lower mass systems, is likely due to using the slope of \citet{ble20} as a prior, a large SZ detected cluster sample that extends to lower masses.

A similar study by \citet{mul19} is composed of low-z, high-$\lambda$ clusters based on Subaru/Suprime-Cam observations.  Taking the already converted scaling relation ($M_{500c}$ to $M_{200c}$) from \citet{kii20}, the prediction has a steeper slope, but agreement in the richness range they fit for ($\lambda > 50$) is good. Extrapolating to lower richness though, agreement is poor in general.  Our sample demonstrates the importance of low mass clusters when studying scaling relations.

Carrying out the same analysis on the high redshift sample produced a completely unconstrained fit.  With less than 30 examples of clusters at $z > 0.4$, all with high richness as well, much more low richness data is required.

\newcommand{\x}{\sigma}
\newcommand{\obs}[1]{\tilde{#1}}
\newcommand{\true}[1]{#1}
\newcommand{\zmin}{0.1\xspace}
\newcommand{\zmax}{0.3\xspace}
\newcommand{\xmin}{497\xspace}
\newcommand{\xmax}{2230\xspace}
\newcommand{\bin}[1]{\Delta #1}
\newcommand{\binlnx}{\bin{\!\ln{\obs{\x}}}_i}
\newcommand{\binx}{\bin{\obs{\x}}_i}
\newcommand{\binz}{\bin{{\obs{z}}}_j}
\newcommand{\codex}{CODEX\xspace}
\newcommand{\xclass}{XCLASS\xspace}
\newcommand{\spiders}{SPIDERS\xspace}
\newcommand{\selectargs}{\true{\mu}, \true{z}, \nu}
\newcommand{\intlim}[2]{\int_{#1}^{#2}}
\newcommand{\intzobs}{\int_{\binz}}
\newcommand{\intzobsi}{\int_{\binzi}}
\newcommand{\intlnxobs}{\int_{\binlnx}}
\newcommand{\intnutrue}{\intlim{-4}{+4}}
\newcommand{\intztrue}{\intlim{0.05}{0.75}}
\newcommand{\lnxlimbase}{\expect{\ln\true{\x}(\true{\mu}, \true{z})}}
\newcommand{\intlnxtrue}{\intlim{\lnxlimbase - 4\sintlnx}{\lnxlimbase + 4 \sintlnx}}
\newcommand{\intmutrue}{\intlim{30.7}{35.3}}
\newcommand{\prob}[1]{P(#1)}
\newcommand{\g}[2]{(#1\,|\,#2)}
\newcommand{\probg}[2]{P\g{#1}{#2}}
\newcommand{\probgg}[3]{P_{#1}\g{#2}{#3}}
\newcommand{\expect}[1]{\langle #1 \rangle}

\section{Cosmology constraints} \label{sec:7}

Because velocity dispersion is a proxy for cluster mass, the evolution of the velocity dispersion function is a useful cosmological tool. The primary cosmological analysis for the \codex component of the \spiders DR16 cluster sample is carried out using optical richness as a mass proxy \citep{IderChitham2020}. Constraints can be derived in an analogous way using velocity dispersion as a dynamical mass proxy. To accomplish this, a modification to the PDF which describes the relation between the observed and true quantities is necessary i.e. replacing $\probg{\ln\obs{\lambda}}{\ln\true{\lambda(\true{\mu},\true{z}}}$ with $\probg{\ln\obs{\sigma}}{\ln\true{\sigma}(\true{\mu},\true{z}})$ in the likelihood function presented in \citep{IderChitham2020}, where $\probg{\ln\obs{\sigma}}{\ln\true{\sigma}(\true{\mu},\true{z}})$ is a normal distribution centred on the value of the natural logarithm of the true velocity dispersion $\ln\true{\sigma}$  \citep{Carlberg1997aa} added to a mean bias value, with a scatter of both statistical and intrinsic origin. Here we define $\mu = \ln M_{\rm 200c}$. 
Based on numerical simulations, \citet{ferragamo2020} found the mean bias value to depend on the radial aperture (through a multiplicative factor $B_{\rm ape}$) and number of galaxies $N_{\rm gal}$ entering the measurement and on the fraction of interlopers. Interlopers are galaxies entering the computation of velocity dispersion, but they are located outside of the virially bound cluster region. Following results of \citet{saro2013}, the fraction of interlopers is estimated as a function of cluster redshift and radial aperture. This fraction sets the value of the multiplicative bias factor $B_{\rm int}$ on the velocity dispersion measurement \citep{ferragamo2020}. An additional 0.98 bias factor is applied, considering that only the most massive galaxies contribute to the measurement \citep{ferragamo2020}. We found this model is consistent with that derived from SPIDERS-like resampling of bright nearby clusters \citep{zha16}.
Combining these four sources of bias and following our notations, we write:
\begin{equation}
    \langle \obs{\sigma} \rangle = 
   \frac{0.98\sigma }{0.9775 +\left(\frac{0.72}{N_{\rm gal}-1}\right)^{1.28}}
   B_{\rm ap}\left( \frac{R_{\rm ap}}{R_{\rm 200c}(\mu,{z})} \right)
   B_{\rm int} \left( \frac{R_{\rm ap}}{R_{\rm 200c}(\mu,{z})}, z \right).
\end{equation}

As for estimating the scatter, we considered the model of \citet{saro2013} (their Eq.~7) which links $N_{\rm gal}$ to the scatter on $\ln(\obs{\sigma}/\sigma)$. This is added in quadrature to the extra contribution from interlopers shown in their Fig.~10, that is a function $S_{int}$ of both cluster mass and radial aperture.  Therefore we write:
\begin{equation}
    Var( \ln \obs{\sigma} ) = \left(-0.037+\frac{1.047}{\sqrt{N_{\rm gal}}}\right)^2 + \left(S_{\rm int}\left(\frac{R_{\rm ap}}{R_{\rm 200c}(\mu, z)}, M_{\rm vir}(\mu, z) \right) \right)^2 ,
\end{equation}
and for simplicity, we define $\delta = \ln \langle \obs{\sigma}  \rangle - \ln\sigma$ and the scatter on  $\ln\obs{\sigma}$ as $\Delta = \sqrt{Var(\ln \obs{\sigma} )}$.

Using now the assumption that $\ln \obs{\sigma}$ is normally distributed around $\ln\true{\sigma}$ with a bias term (equivalently, $\obs{\sigma}$ distribution is log-normal), we write the following expression for the additive bias $b$ on $\ln\obs{\sigma}$:
\begin{equation}
b(R_{\rm ap},  M_{\rm vir}(\mu, z),  N_{\rm gal}) = \langle \ln \obs{\sigma} \rangle - \ln \sigma = \delta - \frac{\Delta^2}{2}.
\end{equation}
The scatter term $\Delta$ due to measurement uncertainties is added in quadrature to an intrinsic scatter to form the resulting scatter on $\ln\obs{\sigma}$:
\begin{equation}
    f = \sqrt{\Delta^2 + \Sigma_0^2}.
\end{equation}

%
%
Presented in section~\ref{sec:6}, velocity dispersion can be linked to the total mass by equation~\ref{eq:sigma-mass}.
This relation describes the mean value of $\sigma$. We consider the scatter term as a part of $\obs{\sigma}$. The discussion above can be summarized as a probability to observe a velocity dispersion, for a given true value of it and a redshift (which defines the quality of the data). We denote this as $P(\ln \obs{\sigma}| \ln \sigma (\mu,z),z)$. 
The relation is controlled by a single number, $N_{\rm gal}$, that we tabulated based on the data and inverted the relation to read
$N_{\rm gal}(\sigma(\mu,z), z)$, a function called by the code, which subsequently defines the bias $b$ and the scatter $f$ of the observed values,

\begin{multline}
   P(\ln \obs{\sigma}|\ln\sigma (\mu,z),z) =\mathcal{N}(\ln \obs{\sigma}, \\
   \ln \sigma + b(R_{\rm ap}, M_{\rm vir}(\mu, z), N_{\rm gal}), \\
   f(R_{\rm ap}, M_{\rm vir}(\mu, z), N_{\rm gal})).
\end{multline}

The error term is done by sampling the data and determining the typical number of members entering a velocity dispersion calculation at a given dispersion and redshift bin $f( \sigma, z)$. The way to estimate the error using an average number of members is given above.

A 10\% sensitivity cut to the sample, proposed by \cite{fin20} to obtain clean X-ray identification, corresponds to
\begin{equation} \label{eqn:cut1}
P^\mathrm{RASS}(I | \obs{\sigma}, z)=\theta (\obs{\sigma} -  375(z/0.15)^{0.38}),
\end{equation}
where $\theta$ denotes a step-function. To avoid incompleteness of SDSS, we introduce an additional cut:
\begin{equation} \label{eqn:cut2}
P^\mathrm{SDSS}(I>0.9 | \obs{\sigma}, z)=\theta(\obs{\sigma}- (372 + 2.76e^{3.3z^2})).
\end{equation}
This equation describes the 90\% completeness of SDSS. Since the cuts are identical between the data and the model, to ensure the cut is applied consistently to both data and the model, we include it in a definition of velocity dispersion bins, $\Delta \ln \obs{\sigma}_i$, to stay above the completeness limits. The CODEX survey function $P_X(I|\mu,z,\nu=0)$ \citep{fin20} is included, where $\nu=0$ notation means that we use the part of the calculation, which corresponds to a zero value of covariance between the velocity dispersion and X-ray luminosity: 

\begin{multline}
\label{eqn:hmf}
\langle N(\Delta \ln \obs{\sigma}_i, \Delta z_j) \rangle = \int_{\Delta z_j} \de z  \frac{\de V}{\de z} \int_{\Delta \ln \obs{\sigma}_i} \de \ln \obs{\sigma}  \\
 \int \de \mu
  P(\ln \obs{\sigma}|\ln \sigma(\mu,z),z) P_X(I | \mu,z,\nu=0) 
   \frac{\de n(\mu, z) }{\de\mu}. 
\end{multline}

\begin{figure}
    \centering
    \includegraphics[width=\columnwidth]{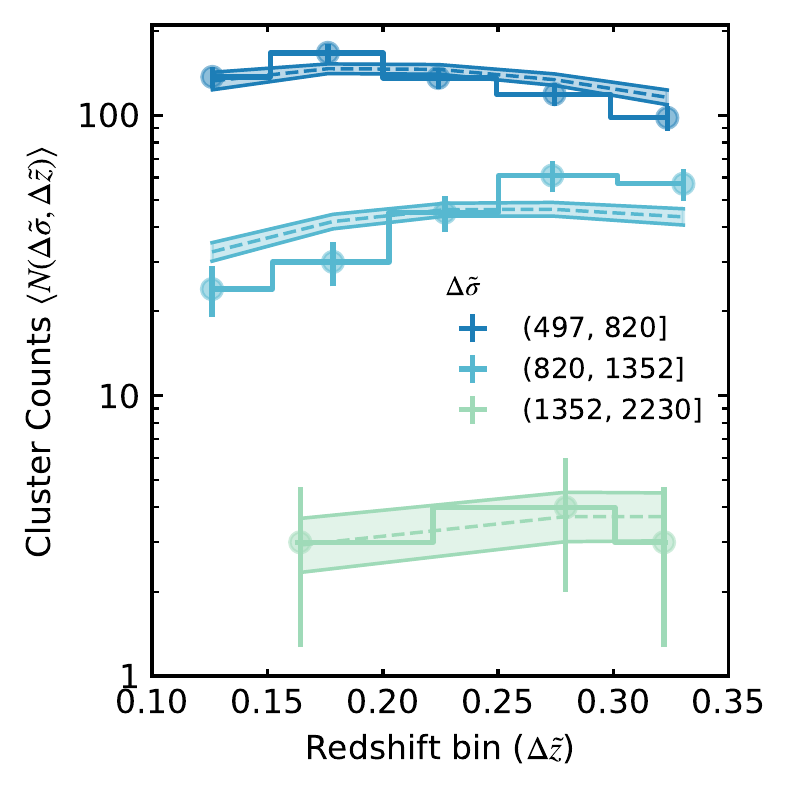}
    \caption{\label{fig:delta_z} The abundance of \spiders clusters as a function in bins ($\Delta$) of observed redshift ($\binz$) and velocity dispersion ($\binx$) where $\obs{z} \in \left[\zmin, \zmax\right)$ and $\obs{\x} \in \left[\xmin, \xmax\right).$
    Systems are selected according to Equation~\ref{eqn:cut1} and Equation~\ref{eqn:cut2}. These cuts on the observable space is accounted for in the cosmological likelihood function. Steps represent the observed data and shaded regions trace the expectation value of the model, centred on the median. The upper and lower limit correspond to the 15\% and 85\% confidence intervals of the MCMC chains.}
   
\end{figure}

The binned data and best fit model are shown in Figure~\ref{fig:delta_z} for three logarithmically distributed bins of velocity dispersion over a redshift range of $[0.1, 0.3)$ with a bin-width of 0.05. The low velocity dispersion bin is dominated by selection effects, yet it provides the strongest constraints on the $z \sim 0.15$ cluster abundance. The two high velocity dispersion bins show the effect of increasing volume, which balances out the evolution of the cluster mass function. The evolution of the highest bin shall be stronger and so, the resulting trend is flatter.

\begin{figure}
    \centering
    \includegraphics[width=\columnwidth]{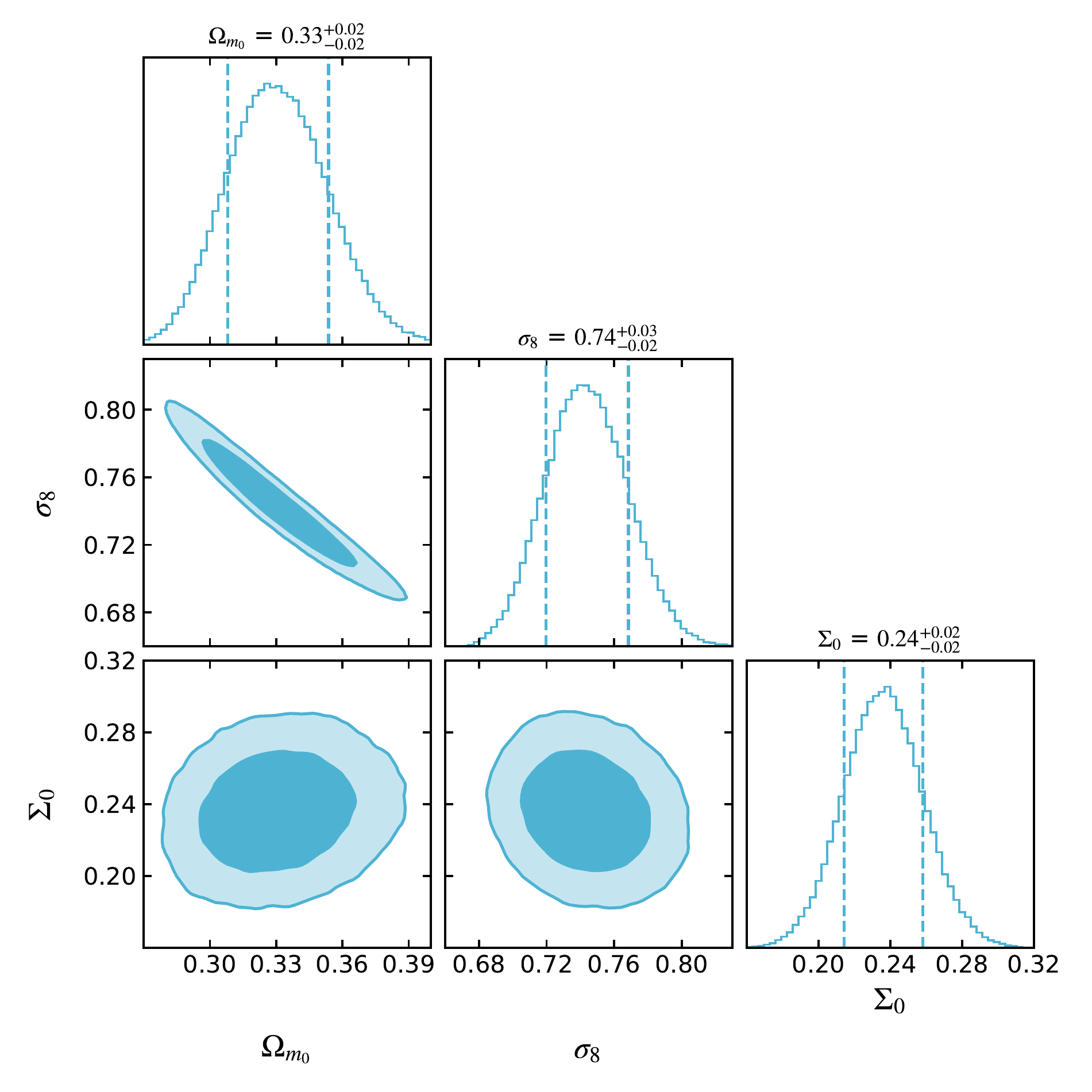}
    \caption{\label{fig:cosmo_results}
    Constraints on cosmological parameters and the intrinsic scatter of the velocity dispersion constraints for \spiders DR16 as. Contours depict the 68\% and 95\% confidence levels where posterior distributions are obtained using the likelihood function described in section~\ref{sec:7}.}
\end{figure}

The corresponding constraints are $\sigma_8=0.74^{+0.03}_{-0.02}$, $\Omega_{m_0}=0.33^{+0.02}_{-0.02}$ and the velocity dispersion of the intrinsic scatter $\Sigma_{0}=0.24^{+0.02}_{-0.02}$ derived from the Markov Chain Monte Carlo (MCMC) process are shown in Figure~\ref{fig:cosmo_results}. These results depend on a large number of fixed parameters which reduce the flexibility of the likelihood, causing a significant underestimation of the apparent uncertainties.  However, our solution still overlaps with other cluster surveys \citep{vik09,boh14,boc19,fin20}.  Cosmology using a non-baryonic tracer of mass is similar to that achieved through calibration of baryonic tracers. Since the need for cluster calibration is removed, the tension to Planck18 cosmology can no longer be explained through deficiencies of mass calibration.  In the calibration of $\sigma$, we rely on numerical simulations \citep{Carlberg1997aa} and not on the empirical calibrations. Since it is these empirical calibrations that have been questioned during the Planck CMB vs. Planck cluster cosmology comparison \citep{pc18}, this argument cannot be used to explain the disagreement of our results with Planck18 cosmology and other explanations shall be sought.  We have estimated, for the first time the intrinsic scatter of $\sigma$.  Compared to simulations, our estimate falls between the higher prediction of \citet{saro2013} and the lower prediction of \citet{mun13}.  The low intrinsic scatter is also comparable to other cluster mass proxies such as X-ray luminosity, X-ray temperature, optical richness and SZ thermal energy \citep[][]{cap19,cap20,mul19,ble20,kii20}

\section{Summary}

We present the largest catalogue of visually confirmed, spectroscopic galaxy clusters.  Herein we describe the entire validation process.  The SPIDERS program successfully followed-up 2740 X-ray sources identified as galaxy clusters detected in RASS.  A total of 33,340 redshifts were used with an improved $\Delta z_{\rm{spec}} /(1+z) = 6 \times 10^{-4}$ compared to $\Delta z_{\rm{phot}} /(1+z) = 5 \times 10^{-2}$.  The overall efficiency of successful validation reached 99\% when $\geq 9$ red-sequence members were observed.  This improves to $\geq 5$ red-sequence members observed at redshifts $z<0.3$.  The number of confirmed members per cluster ranges from 3 to 75, with a median of 10 members.

We fit a scaling relation as $\log \sigma \propto A + B  \log L_{\rm{X}}$ (equation~\ref{eq:sig_lx}) using the BCES method.  The resulting normalization and slope are:
\begin{equation}
\begin{split}
    A = (-5.91 \pm 1.47) \times 10^{-2} \\
    B = (1.29 \pm 0.35) \times 10^{-1}.
\end{split}
\end{equation}
Our relation is in excellent agreement with a previous \citep{cap20} study of dynamical masses using a subset of our sample.

We fit as well a scaling relation as $\log \sigma \propto A + B  \log \lambda_{\rm{Legacy}}$ (equation~\ref{eqn:sig_rich}) using the BCES method as well.  The resulting normalization and slope are:
\begin{equation}
\begin{split}
    A = (-5.36 \pm 0.45) \times 10^{-2} \\
    B = (3.23 \pm 0.57) \times 10^{-1}.
\end{split}
\end{equation}
Comparing to weak lensing (WL) samples \citep[][]{mul19,kii20}, our relation compares well in the high richness range.  Extrapolating to low richness the two WL samples diverge from agreement.  Our sample highlights the importance the large richness range is for constraining scaling relation models.

On a first attempt to constrain cosmology with this sample, velocity dispersion is used as a mass proxy.  The constraints we obtain for as follow:
\begin{equation}
\begin{split}
    \sigma_8=0.74^{+0.03}_{-0.02} \\
    \Omega_{m_0}=0.33^{+0.02}_{-0.02} \\
    \Sigma_{0}=0.24^{+0.02}_{-0.02}.
\end{split}
\end{equation}
Though the errors are underestimated, our solution overlaps with other galaxy cluster surveys.  We find the values of scatter being comparably low to other cluster mass tracers, which makes $\sigma$ a viable cosmology tool and provides good prospects for the eROSITA cluster follow-up program on 4MOST \citep{fin19}. 

\section*{Data Availability}
The data underlying this article are available in the SDSS website at https://www.sdss.org.

\section*{Acknowledgements}

CK, NC, AF and JIC acknowledge the financial support from the visitor and mobility program of the Finnish Centre for Astronomy with ESO (FINCA), funded by the Academy of Finland grant nr 306531.

CK and NC acknowledge this research was supported by the DFG cluster of excellence `Origin and Structure of the Universe' (www.universe-cluster.de).

NC acknowledges travel support from the Institut Francais de Finlande, the Embassy of France in Finland, the French Ministry of Higher Education, Research and Innovation, the Finnish Society of Sciences and Letters and the Finnish Academy of Science and Letters.

AS is supported by the ERC-StG `ClustersXCosmo' grant agreement 716762, and by the FARE-MIUR grant `ClustersXEuclid' R165SBKTMA.

Funding for the Sloan Digital Sky Survey IV has been provided by the Alfred P. Sloan Foundation, the U.S. Department of Energy Office of Science, and the Participating Institutions. SDSS-IV acknowledges
support and resources from the Center for High-Performance Computing at
the University of Utah. The SDSS web site is www.sdss.org.

SDSS-IV is managed by the Astrophysical Research Consortium for the 
Participating Institutions of the SDSS Collaboration including the 
Brazilian Participation Group, the Carnegie Institution for Science, 
Carnegie Mellon University, the Chilean Participation Group, the French Participation Group, Harvard-Smithsonian Center for Astrophysics, 
Instituto de Astrof\'isica de Canarias, The Johns Hopkins University, Kavli Institute for the Physics and Mathematics of the Universe (IPMU) / 
University of Tokyo, the Korean Participation Group, Lawrence Berkeley National Laboratory, 
Leibniz Institut f\"ur Astrophysik Potsdam (AIP),  
Max-Planck-Institut f\"ur Astronomie (MPIA Heidelberg), 
Max-Planck-Institut f\"ur Astrophysik (MPA Garching), 
Max-Planck-Institut f\"ur Extraterrestrische Physik (MPE), 
National Astronomical Observatories of China, New Mexico State University, 
New York University, University of Notre Dame, 
Observat\'ario Nacional / MCTI, The Ohio State University, 
Pennsylvania State University, Shanghai Astronomical Observatory, 
United Kingdom Participation Group,
Universidad Nacional Aut\'onoma de M\'exico, University of Arizona, 
University of Colorado Boulder, University of Oxford, University of Portsmouth, 
University of Utah, University of Virginia, University of Washington, University of Wisconsin, 
Vanderbilt University, and Yale University.






\bibliographystyle{mnras}
\bibliography{myref} 

 




\bsp	
\label{lastpage}
\end{document}